%% file: ERNewScale.tex
\pdfoutput=1

\documentclass[aps,prd,twocolumn,showpacs,superscriptaddress,groupedaddress,nofootinbib]{revtex4}  
\usepackage{graphicx}  
\usepackage{dcolumn}   
\usepackage{bm}        
\usepackage{amssymb}   
\usepackage{xcolor}

\usepackage[perpage,stable]{footmisc}
\usepackage[none]{hyphenat}

\begin{document}
 


\title{Scintillation and Ionization Responses of Liquid Xenon to Low Energy Electronic and Nuclear Recoils at Drift Fields from 236\,V/cm to 3.93\,kV/cm}

\input author_list.tex       
\date{\today}

\begin{abstract}
We present new measurements of the scintillation and ionization yields in liquid xenon for low energy electronic (about 3--7\,keV$_{ee}$) and nuclear recoils (about 8--20\,keV$_{nr}$) at different drift fields from 236\,V/cm to 3.93\,kV/cm, using a three-dimensional sensitive liquid xenon time projection chamber with high energy and position resolutions. 
Our measurement of signal responses to nuclear recoils agrees with predictions from the NEST model. 
However, our measured ionization (scintillation) yields for electronic recoils are consistently higher (lower) than those from the NEST model by about 5\,e$^-$/keV$_{ee}$ (ph/keV$_{ee}$) at all scanned drift fields. New recombination parameters based on the Thomas-Imel box model are derived from our data. Given the lack of precise measurement of scintillation and ionization yields for low energy electronic recoils in liquid xenon previously, our new measurement provides so far the best available data covering low energy region at different drift fields for liquid xenon detectors relevant to dark matter searches.

\end{abstract}

\maketitle

\section{\label{sec:introduction} Introduction}
Particle detection technology based on liquid xenon (LXe) has been developed extensively in the last decade, thanks to the advance and promise of the direct dark matter experiments~\cite{DAMA, ZEPLIN-II, ZEPLIN-III, XENON100, LUX, PandaX, XMASS}. 
Among these, the dual phase xenon technique has made significant progress due to its capability for background identification and suppression, and scalability to a large ton-scale target mass. 
Such a technique allows the detection of low energy electronic and nuclear recoils down to sub-keV with both scintillation and ionization signals. 
Understanding the response of LXe to low energy events becomes increasingly important in order to properly assess the background responses in the detector and to precisely extract the dark matter parameters with a positive detection.

In the last few years, low energy nuclear recoils (NRs) were measured extensively either by tagging elastically scattered neutrons from a fixed energy neutron source, e.g. a DD generator \cite{Aprile:05, Manzur:10, Plante:11}, or by modelling the response and comparing with a neutron source, usually $^{252}$Cf and AmBe, with a spread of neutron energy at the MeV level~\cite{Sorensen, Weber, ZEPLINIII, ZEPLIN-III_NR_Yield}. 
Such measurements have yielded better understanding of the scintillation and ionization properties in LXe for NRs below 10\,keV$_{nr}$
(keV$_{nr}$ denotes the nuclear recoil energy, while keV$_{ee}$ denotes the electron equivalent energy.).
The uncertainty of WIMP detection sensitivity was being reduced continuously, especially for low mass WIMPs below 10\,GeV/c$^2$, due to a better understanding of such quantities.

On the other hand, the scintillation and ionization of low energy electron recoils (ERs) were not well measured, partly due to the difficulty to introduce such a low energy ERs in the LXe target. 
The Columbia and Zurich groups~\cite{Columbia, Zurich} measured the Compton scattered electrons by tagging the scattered gammas at a certain angle.
However, these measurements are limited to scintillation yield with large uncertainties. Recently, an attempt to measure ionization yield at 2.82\,keV$_{ee}$ has been realised by using $^{37}$Ar source doped in LXe~\cite{LowERIonizationMeasurement}. 
So far, that is the only measurement of ionization yield below 10\,keV$_{ee}$ and it's only at one given field (3.75\,kV/cm), and also with large uncertainty. 

Understanding LXe's response to low energy ERs are not only important for fully understanding the background for which the ERs are the dominant contribution so far~\cite{LUX}, but also relevant to extracting information from dark matter candidates that produce ERs, such as from the axioelectric effect~\cite{XE100:axion}. 
Several dark matter detectors based on LXe are operated at different electric fields, with different configurations of light and charge detection. 
For example, LUX detector~\cite{LUX} is operated at a relatively low drift field (180\,V/cm), while the ZEPLIN-III~\cite{ZEPLINIII} detector was operated at a high drift field (3.9\,kV/cm). 
Thus a precise measurement of the scintillation and ionization in LXe for low energy ERs at different fields become extremely demanding. 
Also the understanding of the low energy ERs can provide a basis for the background modeling of the solar neutrino in the next-generation large-scale LXe dark matter detectors~\cite{LZ, LZ2}.

Here we report the measurement of ionization and scintillation yield using Compton scattered low energy electrons in LXe at different fields. 
Unlike the measurement performed at~\cite{Columbia, Zurich}, we did not use tagged Compton gammas for a fixed energy ER. 
Instead, we took data for both ionization and scintillation for Compton electrons at all different energies, and extract the ionization and scintillation yields based on the Thomas-Imel recombination model~\cite{BoxModel} (energy below about 7\,keV$_{ee}$). 
By comparing to the simulation of the signal response, we found our measured results are quite different from those in the NEST model~\cite{NEST, NESTWeb, NEST-1.0} for ERs, while the NRs give consistent results.
We also provide the model-independent photon yields for low energy ERs with energy from about 3 to 20\,keV$_{ee}$, based on a combined energy scale.

\section{Experimental Setup}

The measurement was performed in a two-phase LXe time projection chamber (TPC), with four Hamamatsu R8520 photomultiplier tubes (PMTs) on the top and one R11410 PMT on the bottom viewing a 1-cm thick LXe target, allowing simultaneous measurement of both the scintillation (S1) and ionization (S2) signals.
The X-Y positions of events are reconstructed through the S2 hit patterns on the four top PMTs.
More details of the setup was described in~\cite{resolutionpaper}. Low energy ERs in LXe are obtained using an external $^{137}$Cs source. 662\,keV gamma rays from the source also provide energy calibration and stability monitoring. A $^{252}$Cf neutron source is used to produce low energy NRs in the detector. 
 
\begin{figure}[htp]
 \center
 \includegraphics[width=8cm,height=6cm]{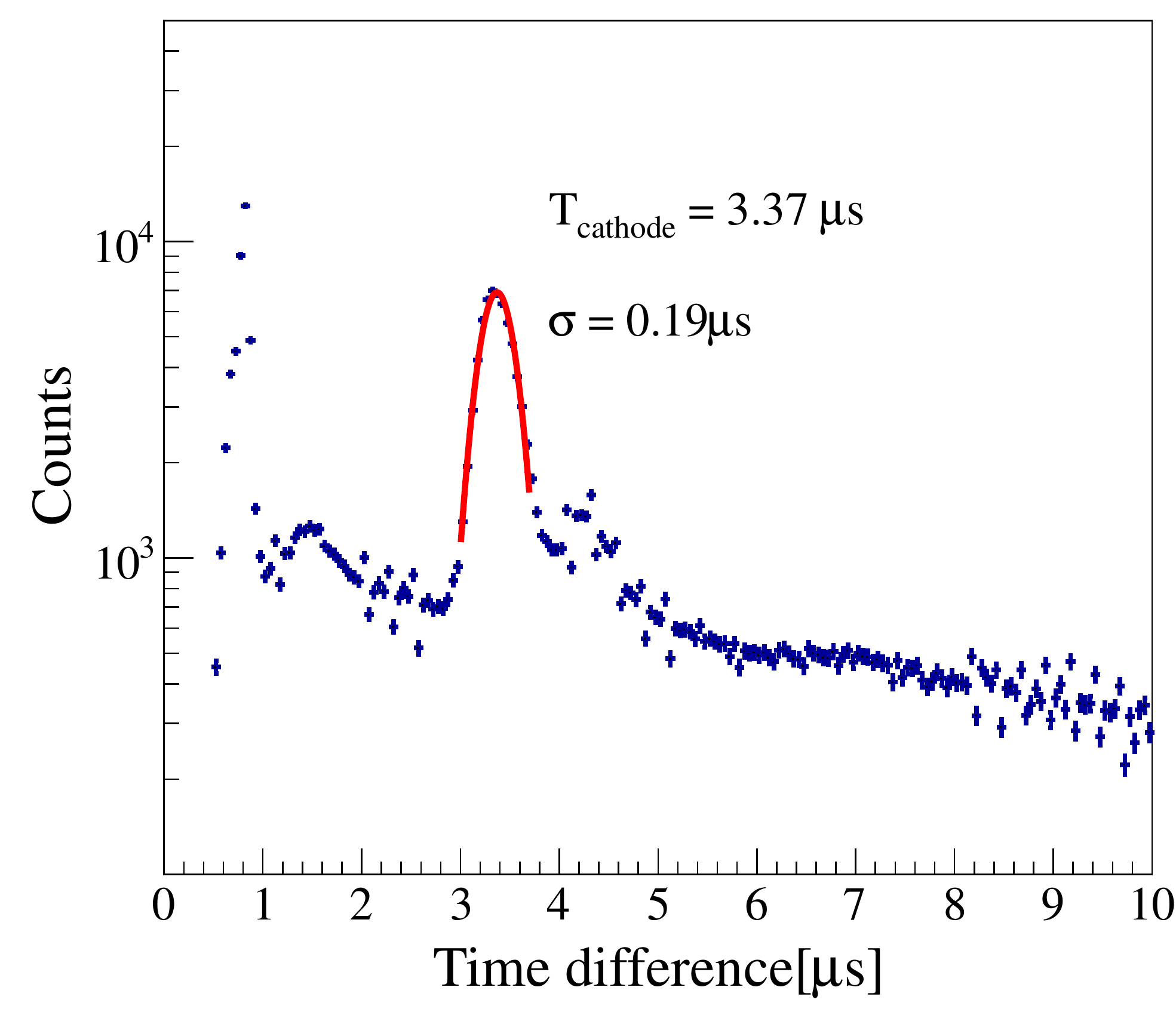}
 \caption{\small Distribution of time differences between a main S2 and an after-pulse S2 in the special high field run. The peak corresponds to the drift time of photoelectrons emitted by the cathode. The mean drift time from the cathode to the liquid surface is 3.37$\pm$0.19\,$\mu$s.}
 \label{fig:S2APTime}
\end{figure}

During the operation, the anode electrode was connected to ground. 
The gate grid was fixed at --4\,kV, providing sufficient extraction and gas amplification fields of about 11\,kV/cm (with the liquid level of about 2.9\,mm)  for the electron emission into the gas phase. 
The cathode was adjusted accordingly to provide nominal drift fields from 200\,V/cm to 2\,kV/cm (correspond to 236\,V/cm to 1.92\,kV/cm according to~\cite{resolutionpaper}) across the 1-cm drift region. 
Due to the limitation of electron transmission through the gate grid, we conducted a special high drift field run by lowering the liquid surface below the gate grid. During such a run, the gate grid was connected to ground and the cathode was set at --5\,kV. Using the mean drift time (Fig.~\ref{fig:S2APTime}) of  photoelectrons emitted from the cathode in the form of S2 after-pulses~\cite{SingleElectron,ZEPLINII_SingleElectron,ZEPLINIII_SingleElectron} and the electron drift velocities in LXe~\cite{DriftVelocity}, we calculate the total drift length and the drift field in LXe to be 8.1$\pm$0.5\,mm and 3.93$\pm$0.15\,kV/cm, respectively, based on the field simulation using the finite element analysis software package COMSOL~\cite{COMSOL}.

The low energy events relevant to our study are below 7\,keV$_{ee}$, while the energy calibration is based on the 662\,keV$_{ee}$ total absorption events. The low energy data and the calibration data were taken with different PMT gains to avoid saturation for the calibration events. The calibration data were taken with the gains of four top R8520 PMTs at 2$\times$10$^6$ and the bottom R11410 PMT at 2.5$\times$10$^4$. For the low energy data, the gain of the bottom R11410 PMT was at 4$\times$10$^6$. 
All S1 and S2 signals are calculated in unit of 	photoelectron (PE) by dividing the signal outputs by the PMT gains.
The signal from bottom PMT was split into two channels, one for S1 with an external x8 amplifier (CAEN N979B) and another for S2 directly, and fed into flash analog-to-digital converter (FADC) digitizers. S2 from the bottom PMT only was used in the analysis due to the high collection efficiency and resolution of the R11410 PMT. S2 from the bottom PMT was used for trigger and a trigger threshold of 58\,PE, corresponding to about 1.8 extracted electrons, was obtained. 
In addition, a high-energy veto with the upperlimit of about 20000\,PE on the S2 signal was implemented in the trigger of the low-energy recoil measurements in order to assure a good linearity of the S2 signal.

The gains of PMTs used in our study were measured every week to ensure PMT stability. The PMT gain was first measured at a high reference voltage using a weak LED light generating single photoelectrons. Then the LED light was increased and signal dependence on the PMT voltage is measured, covering the entire range of gains used in our measurement. 
During the gain calibration, the 	rate of LED pulses was set to about 100\,Hz. 
The gain dependence on the PMT voltage can be described by the following equation. 
\begin{equation}
 G(V) = G(V_{ref}) \bigg( \frac{V}{V_{ref}} \bigg)^{\kappa}, \label{eq:PMTgain_convenient}
\end{equation}
where $V$ and $G(V)$ represent the PMT voltage and the corresponding gain. $V_{ref}$ is the reference voltage, which is 800\,V for R8520 PMTs and 1500\,V for R11410 PMT. For the bottom R11410 PMT, the combined calibration data during the six weeks of operation show an averaged $G(V_{ref})$ of 3.46$\times$10$^6$ and $\kappa$ of 8.33, with standard deviations of 0.11$\times$10$^6$ and 0.09, respectively.
The gain variation due to different event rate in our measurement shall be small ($<$5\%)~\cite{GainAsRate}. 
Also the quantum efficiency (QE) of R11410 PMT shall not depend on the bias voltage according to Hamamatsu corporation.

\section{Detector Calibration}

The measurements of the low energy ERs and NRs were taken within about two months (Nov. 2013--Jan. 2014), during which daily gamma calibrations using the $^{137}$Cs source at a drift field of 987\,V/cm were carried out to monitor the stability of S1 and S2 signals.  
The event rate in the daily monitoring data is about 400\,Hz (with about 30Hz of the triggers due to the background).
Events in the central region with reconstructed radius less than 10\,mm and drift time between 3.5--8.5\,mm  were selected to reject regions with bad field uniformity.  
We fit the S1 and S2 spectra by an exponential function plus a Gaussian, to obtain the S1 and S2 yields for the 662\,keV gamma rays. 

\begin{figure}[htp]
 \center
 \includegraphics[width=8.5cm, height=6cm]{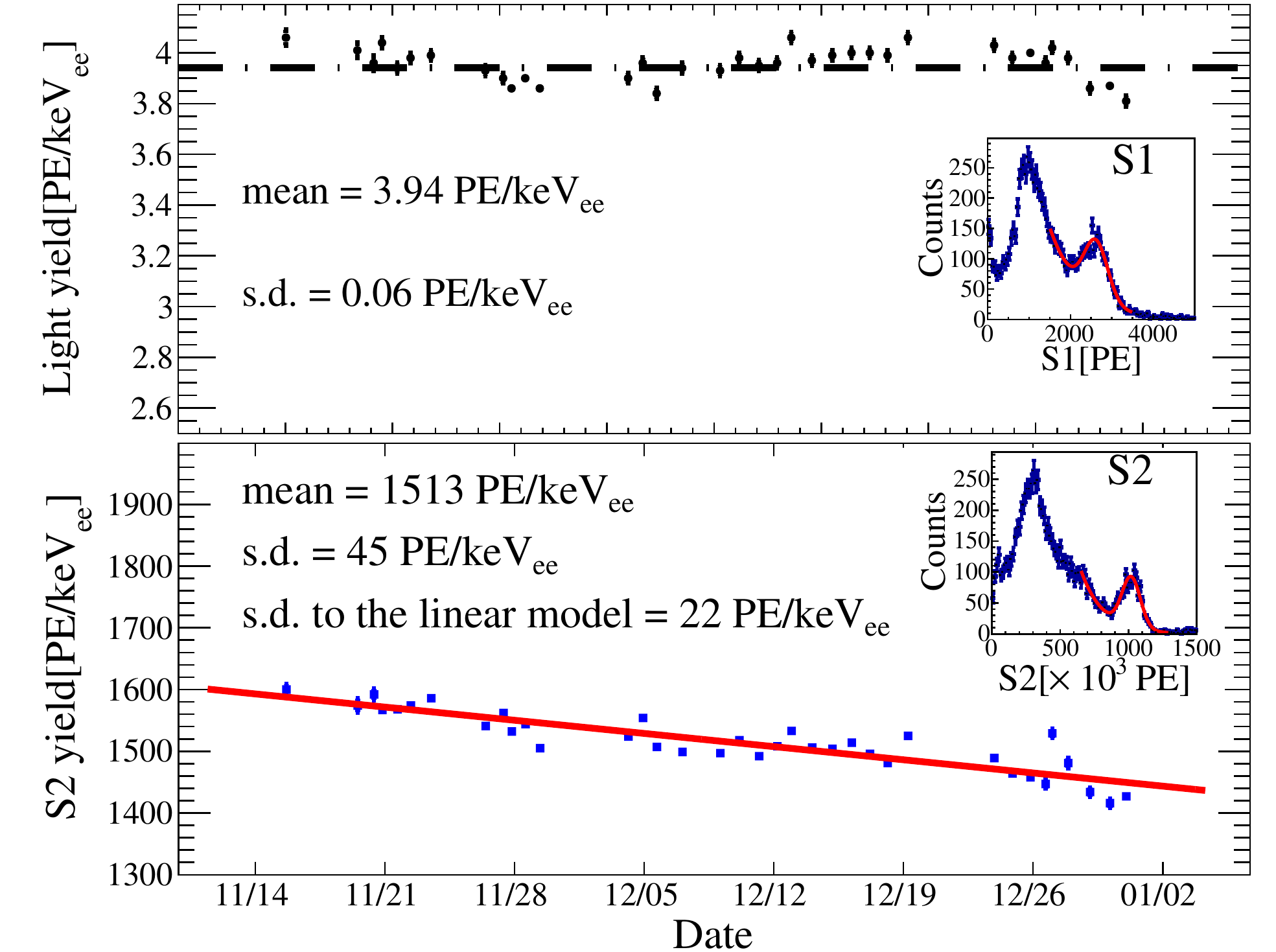}
 \caption{\small The evolution of the S1 light yield (upper) and S2 yield (lower) for 662\,keV gamma rays under a drift field of 987\,V/cm. Typical S1 and S2 spectra, with fits using an exponential function plus a Gaussian, are shown in the insets respectively. The decreasing of the S2 yield over time is modelled by a linear function with a decreasing rate of 3.1$\pm$0.2\,PE/keV$_{ee}$/day with the S2 yield at the beginning of the measurement (15$^{th}$ Nov.) being 1588\,PE/keV$_{ee}$. }
 \label{fig:Monitoring}
\end{figure}

Fig.~\ref{fig:Monitoring} shows the time stability of S1 and S2 yields during the two months of operation. The S1 light yield is stable during the whole period with an average value of 3.94\,PE/keV$_{ee}$ and a standard deviation (s.d.) of 0.06\,PE/keV$_{ee}$. We observe a decrease of S2 yield over time, which is fitted by a linear function with an average value of 1514$\pm$22\,PE/keV$_{ee}$ and a decreasing constant of 3.1$\pm$0.2\,PE/keV$_{ee}$/day. 
The time dependence of the S2 yields is caused by a micro leak in the system gradually lowering the liquid level, reducing the gas field. 
S2 signals of the low energy measurements are corrected for such a time dependence according to the linear fit shown in Fig.~\ref{fig:Monitoring}


The detected S1 and S2 signals can be written as S1 = PDE$\cdot$ N$_\gamma$ and S2 = EAF$\cdot$ N$_e$, where PDE and EAF represent the photon detection efficiency and electron amplification factor, respectively. 
PDE is the product of the light collection efficiency, which is related to the detector geometry, and the quantum efficiency of the PMTs.
EAF is the product of electron extraction efficiency and the gas gain. N$_\gamma$ and N$_e$ are the number of scintillation photons and drifting electrons after the electron-ion recombination process, thus PDE and EAF are independent of drift fields. 
The PDE is mainly relevant to the liquid level 	because of the total reflection of the scintillation light on the liquid-gas surface. The EAF mostly depends on the gas field strength and the thickness of the gas gap.
The photon and electron yield at various drift fields were measured extensively before and their values at 987\,V/cm are 25.43$\pm$1.02\,ph/keV$_{ee}$ and 47.56$\pm$1.90\,e$^-$/keV$_{ee}$, respectively, from NEST v0.98~\cite{NEST}. Based on these values, we obtain a PDE of 15.5$\pm$0.2\% and an EAF of 31.8$\pm$0.5\,PE/e$^-$ from our calibration data. 
The systematic uncertainties for the PDE and EAF are estimated to be $\pm$1.3\% and $\pm$2.6\,PE/e$^-$, respectively, which take into account the global uncertainty of 4\% for the NEST prediction, the S2 yield uncertainty of 1.2\% induced by 20\% electron lifetime variation during the operation, and the gain difference uncertainty of 7.2\% between the monitoring $^{137}$Cs calibration and the low-energy recoil measurements.

For the special run with drift field at 3.93\,kV/cm, because the liquid surface was adjusted below the gate grid, the PDE and EAF are different from the normal runs.  The $^{137}$Cs calibration during the special run gives an average S1 light yield of 3.99$\pm$0.03\,PE/keV$_{ee}$ and an S2 yield of 1053$\pm$5\,PE/keV$_{ee}$. 
The photon yield and electron yield at 3.93$\pm$0.15\,kV/cm for 662\,keV gamma rays are 19.99$\pm$0.81\,ph/keV$_{ee}$ and 52.67$\pm$2.11\,e$^-$/keV$_{ee}$, respectively~\cite{NESTHighFieldYield}.
This leads to a PDE of 20.0$\pm$1.7\% and an EAF of 20.0$\pm$1.7\,PE/e$^-$. 
The S2 signal's time dependence is negligible because it took only two days for the special run and the detector is stable within this time scale.

\section{Results and Discussion}

\begin{figure*}[htp]
 \center
 \includegraphics[width=8cm,height=6cm]{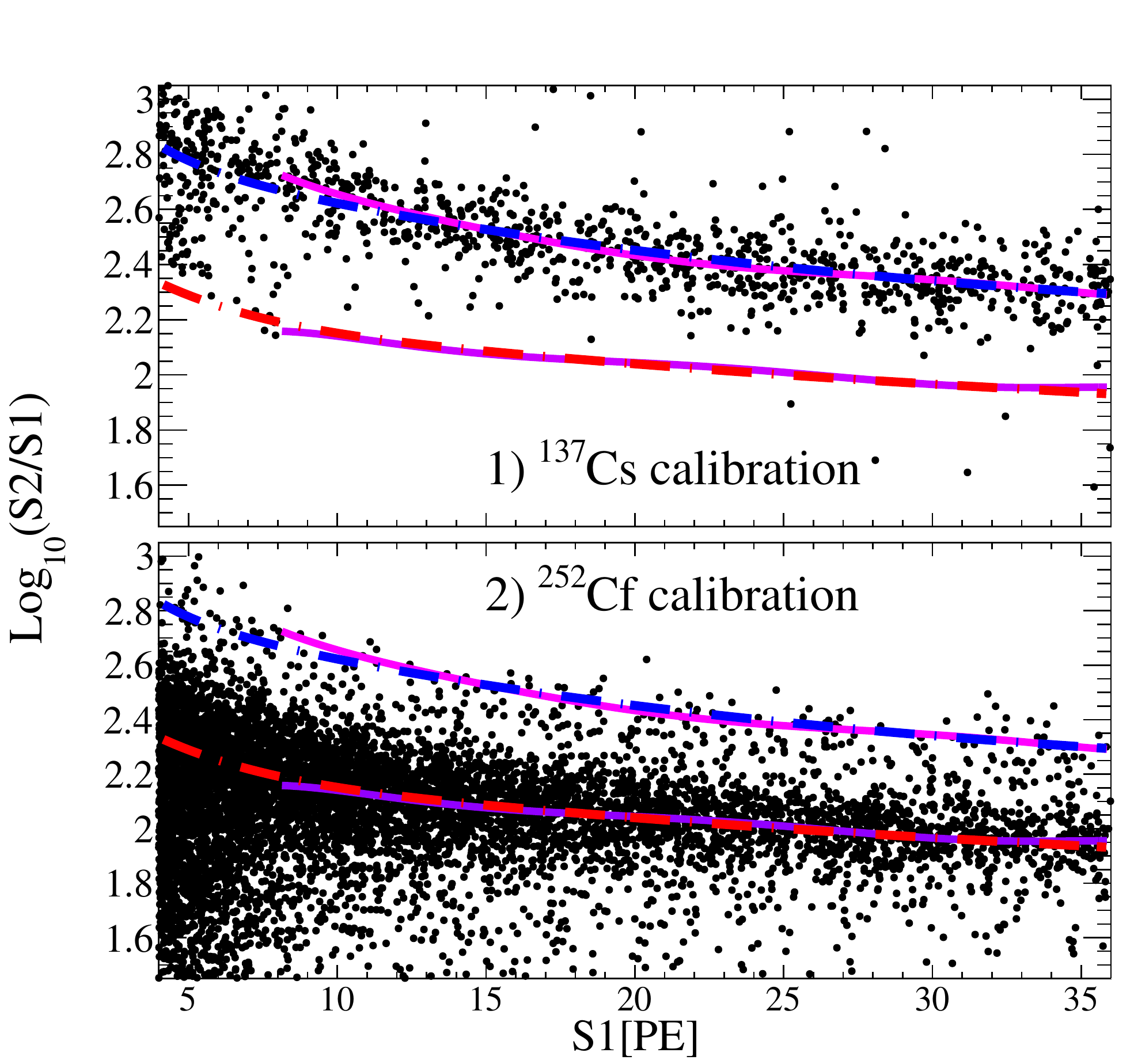} 
 \includegraphics[width=8cm,height=6cm]{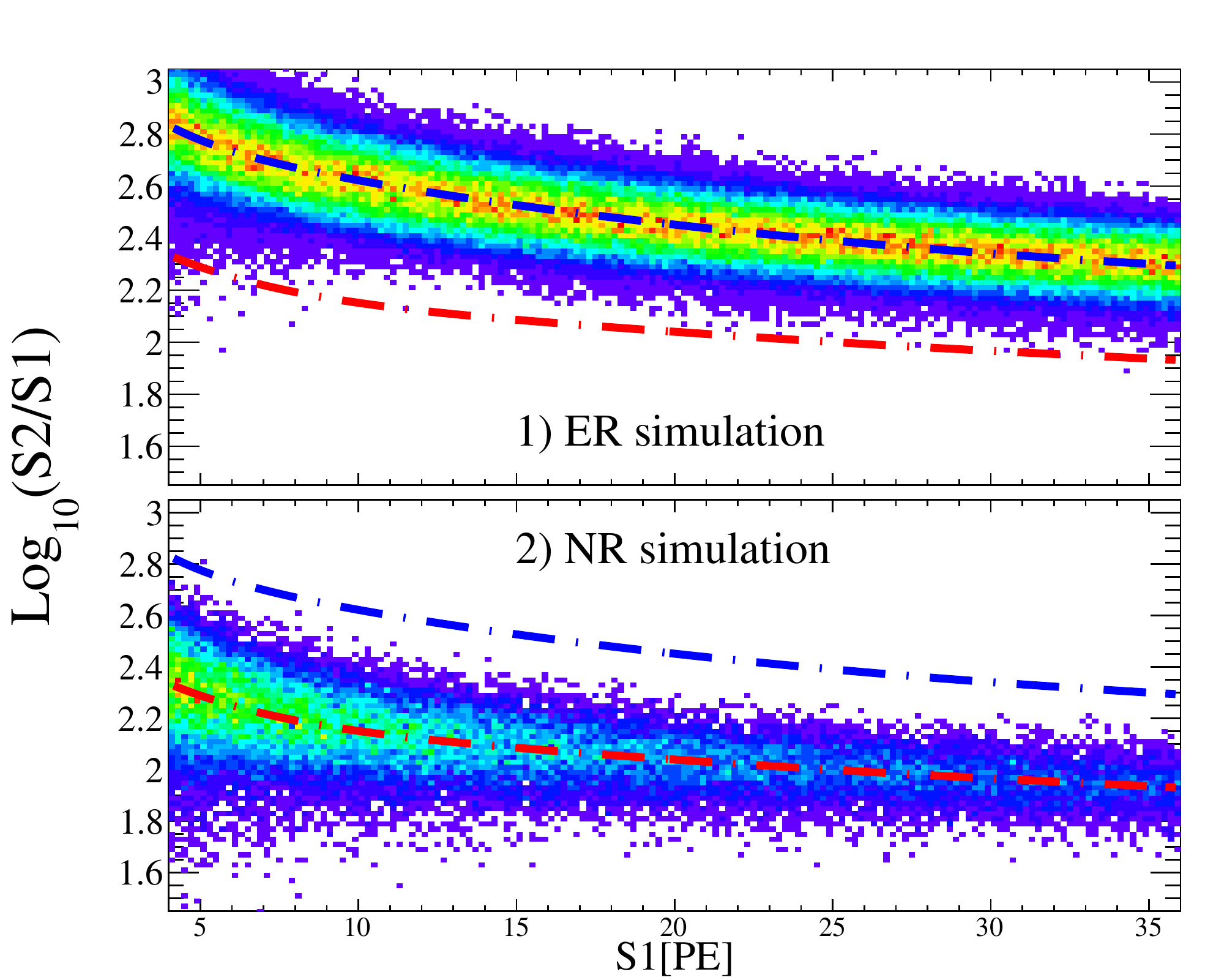}
 \caption{\small The measured (left) low energy ER and NR bands under a field of 236\,V/cm, along with the bands from simulation (right) taking into account the detection efficiencies (PDE and EAF) and statistical smearing effect. 
The magenta (blue) and violet (red) solid (dashed) lines are the means of the ER and NR bands, respectively, in data (MC).
The input of scintillation and ionization yields to the simulation for the NRs is based on the NEST V1.0~\cite{NESTWeb, NEST-1.0}. 
The input for the ERs is based on a $\chi^2$ analysis as discussed in the text.}
 \label{fig:ERNRMeansComparison}
\end{figure*}

Low energy ER data were taken at different drift fields from 236\,V/cm to 3.93\,kV/cm with a low-energy event rate of about 30\,Hz.
To accumulate enough statistics, each measurement was taken for about 24 hours. We also took the NR data at these drift fields with an event rate of about 10\,Hz. 
Single scatter events in the same fiducial volume as for the calibration data are selected for the analysis. As an example, the low energy ER and NR bands at a drift field of 236\,V/cm are shown in Fig.~\ref{fig:ERNRMeansComparison}.

\begin{figure}[htp]
\center
\includegraphics[width=8cm, height=6cm]{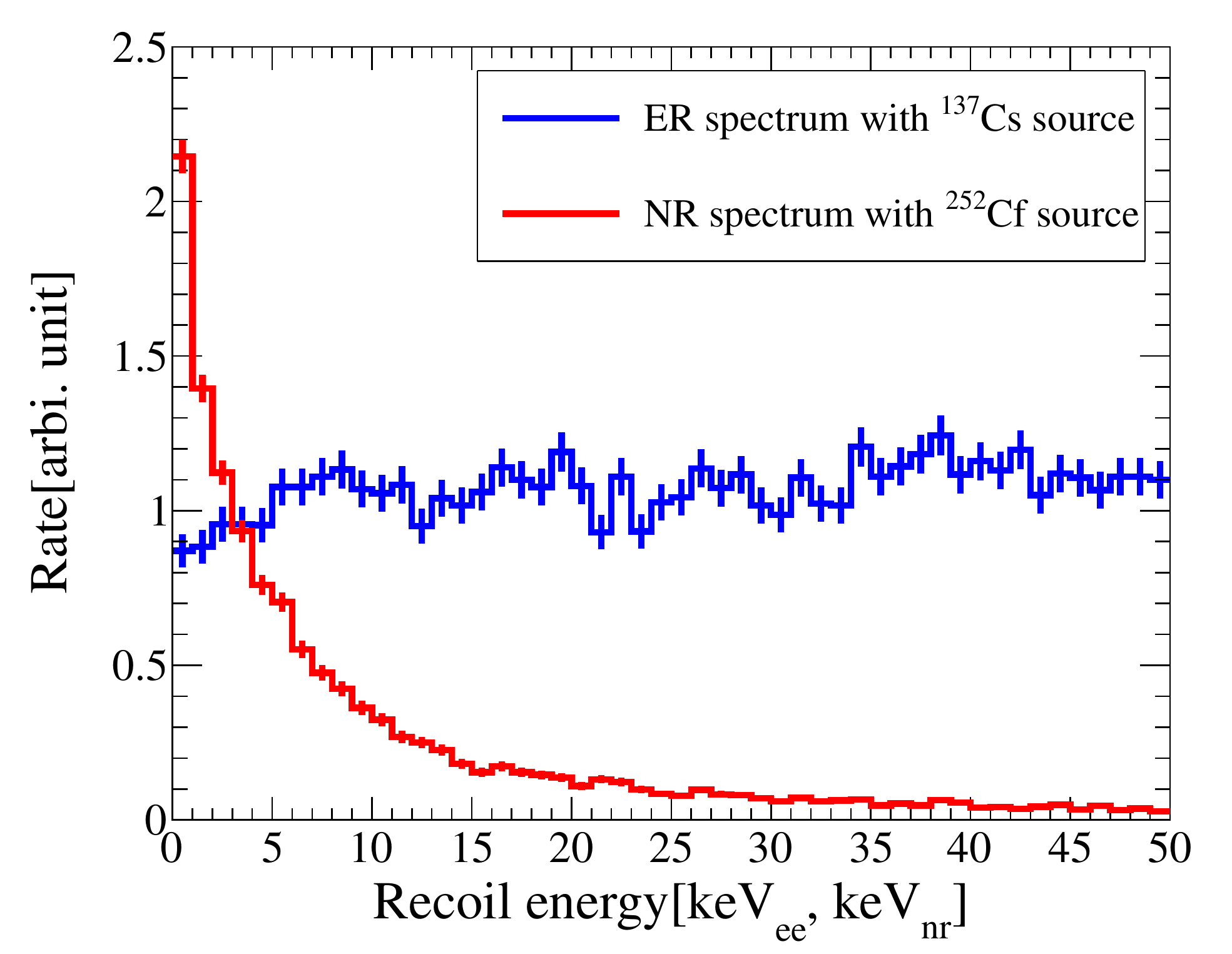}
\caption{\small The ER (blue) and NR (red) spectra obtained through the Geant4 simulation. 
The spectra with energy below 20\,keV$_{ee}$ and 50\,keV$_{nr}$ are used in the signal band simulations for the ER and NR, respectively.
}
\label{fig:G4ERNRSpec}
\end{figure}

\begin{figure}[htp]
 \center
 \includegraphics[width=8cm,height=5.6cm]{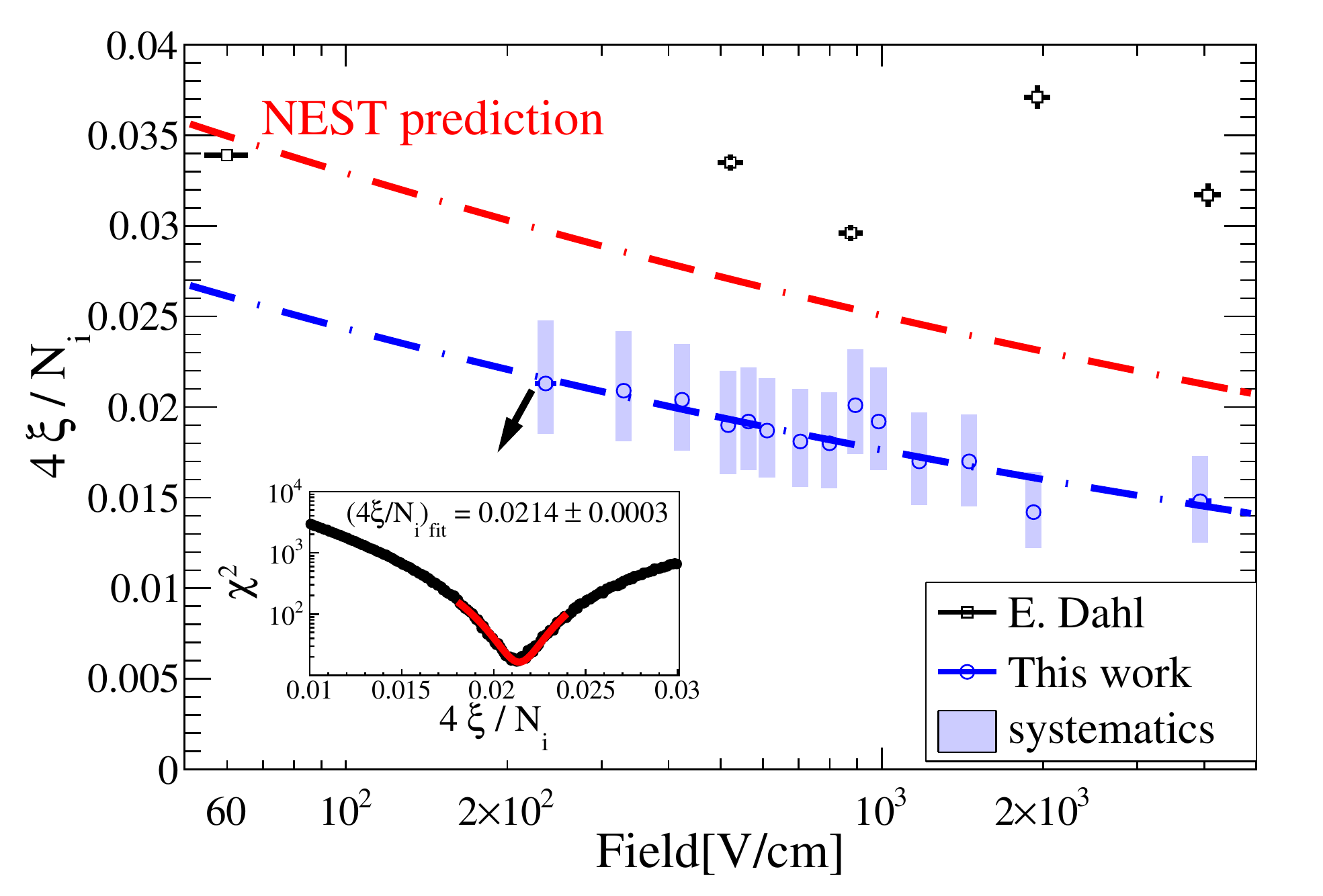}
 \caption{\small The best-fit values of 4$\xi/N_i$ obtained by comparing Monte Carlo (MC) and data for ER. The inset shows the $\chi^2$ of the scanned 4$\xi/N_i$ to the ER means in 236\,V/cm data, which obtains a minimum $\chi^2$=16 (number of degrees of freedom = 7) at 4$\xi/N_i$=0.0214 with a statistical uncertainty of 0.0003 (with $\Delta \chi^2$=1). The red dashed line represents the prediction by NEST. The black rectangles are from E. Dahl~\cite{DahlThesis} (ER only data). The blue shadows represent the systematic uncertainties, which are induced by the uncertainties of the PDE and EAF. A fit through our data points gives $4\xi/N_i$ = 0.046$^{(+0.007)}_{(-0.006)}$ E$^{-0.140}$.
The index of -0.14 is noticeably close to the calculated value of -0.1 by Eric Dahl~\cite{DahlThesis} and the NEST value of -0.11~\cite{NEST}.
}
 \label{fig:ERBoxParameter}
\end{figure}

\begin{figure*}[htp]
 \center
 \includegraphics[width=5cm,height=3.75cm]{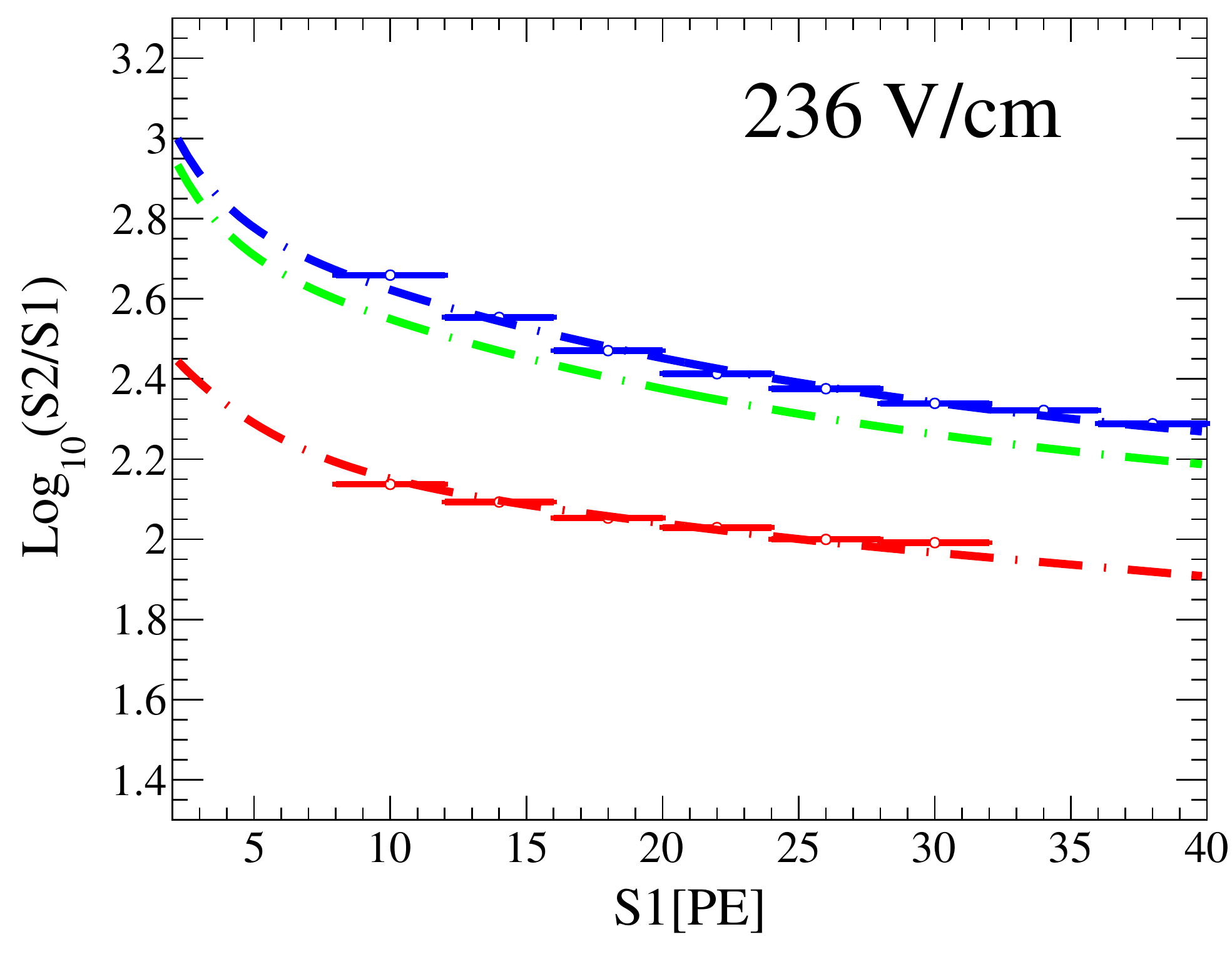}
 \includegraphics[width=5cm,height=3.75cm]{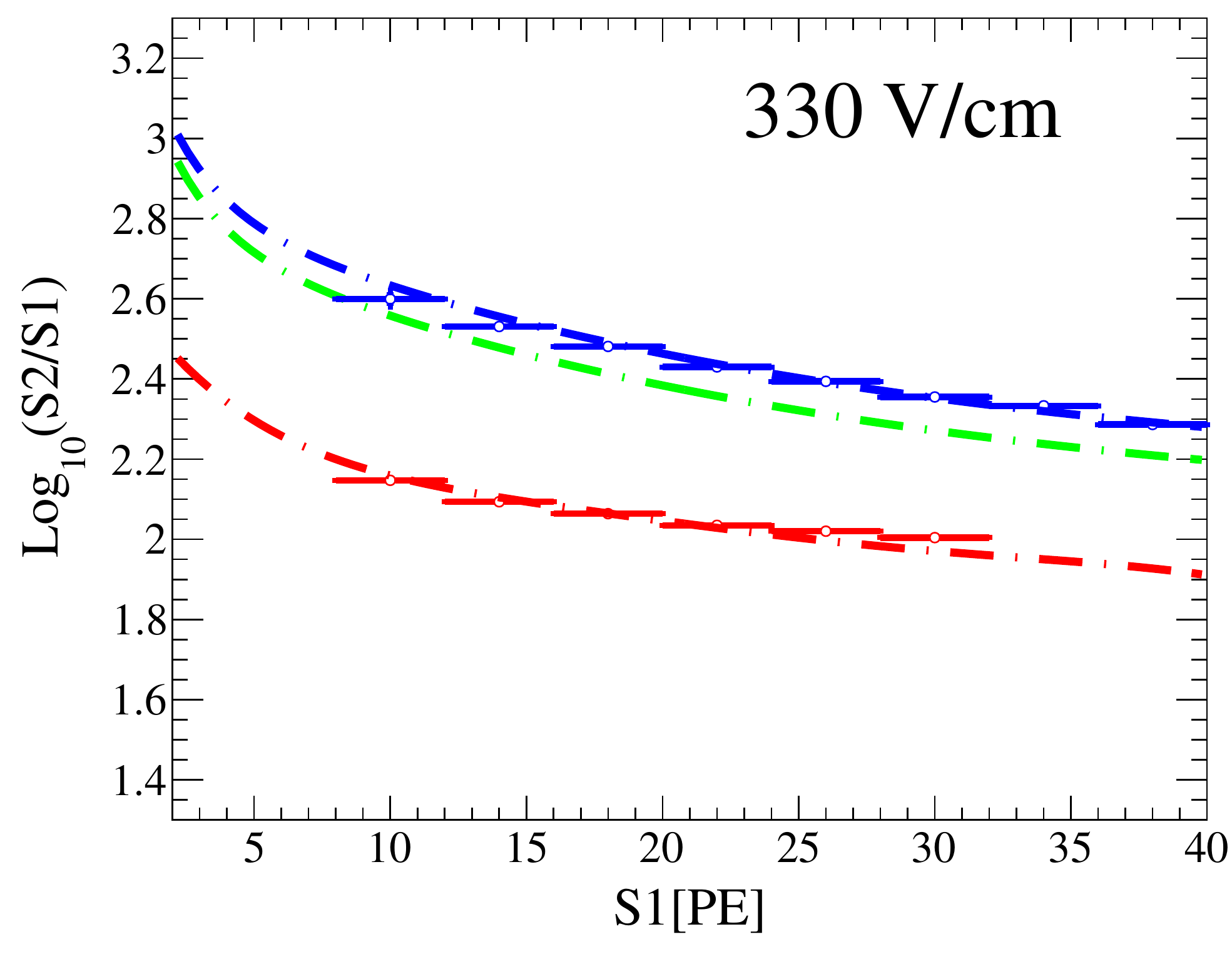}
 \includegraphics[width=5cm,height=3.75cm]{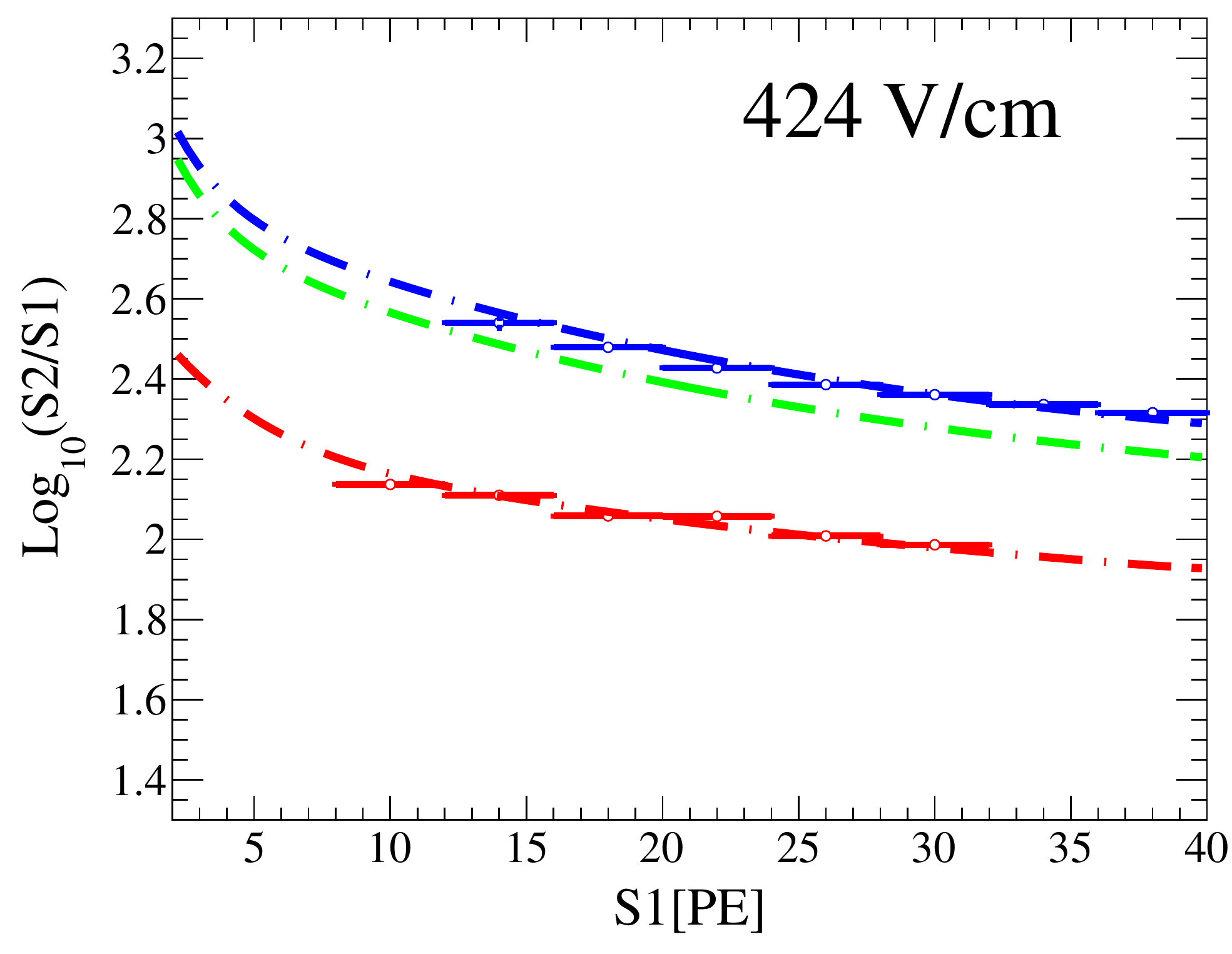}
 
 \includegraphics[width=5cm,height=3.75cm]{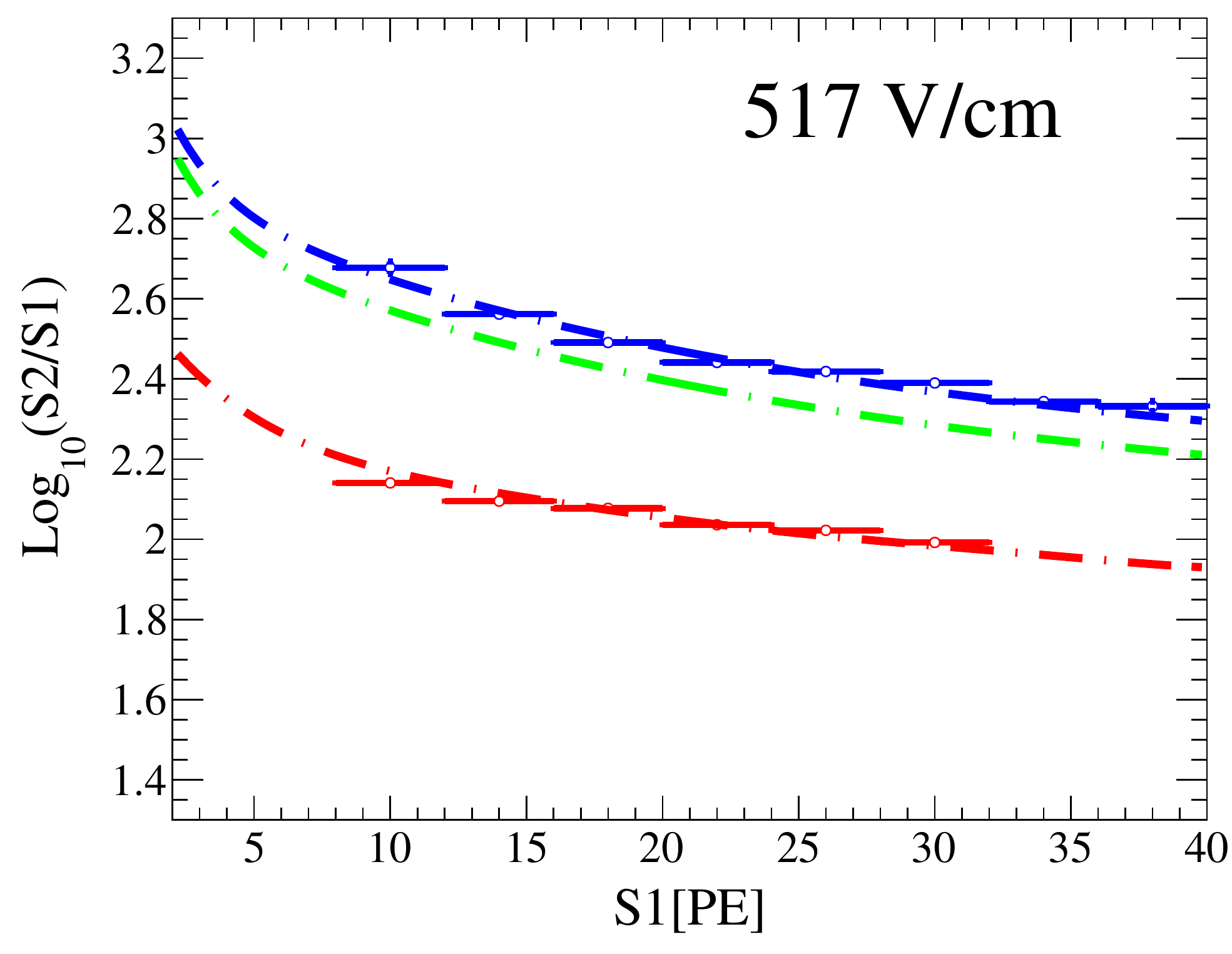}
 \includegraphics[width=5cm,height=3.75cm]{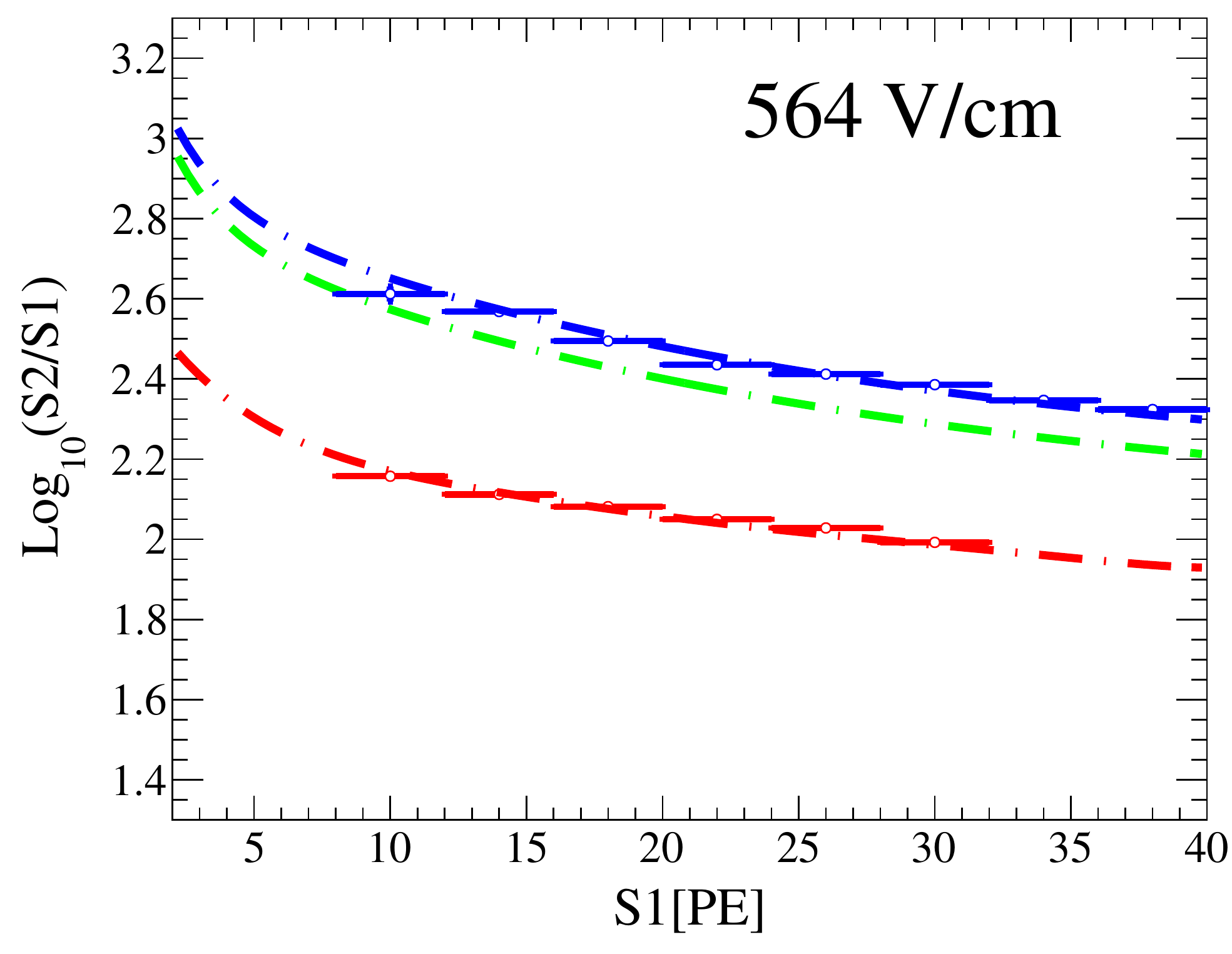}
 \includegraphics[width=5cm,height=3.75cm]{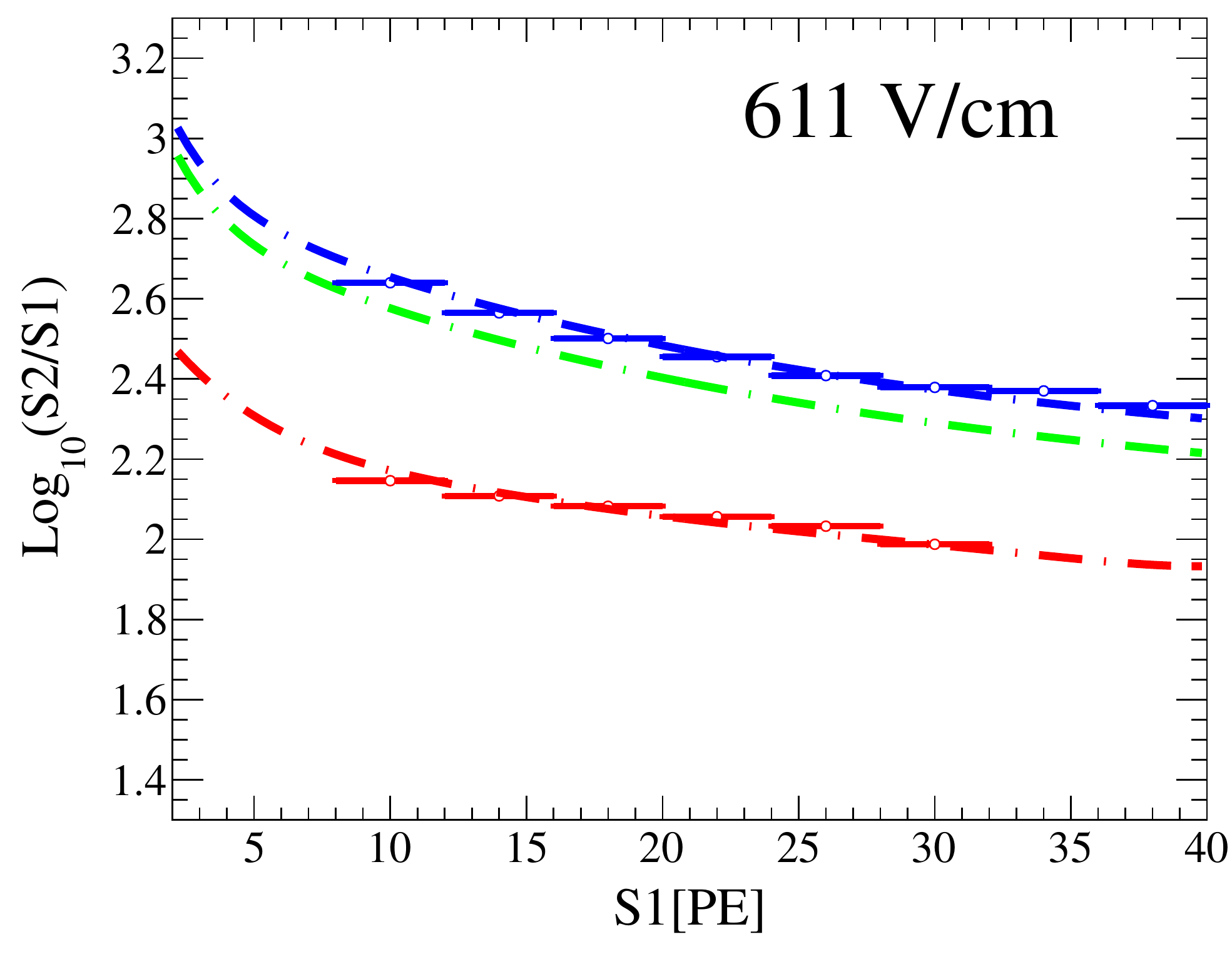}
 
 \includegraphics[width=5cm,height=3.75cm]{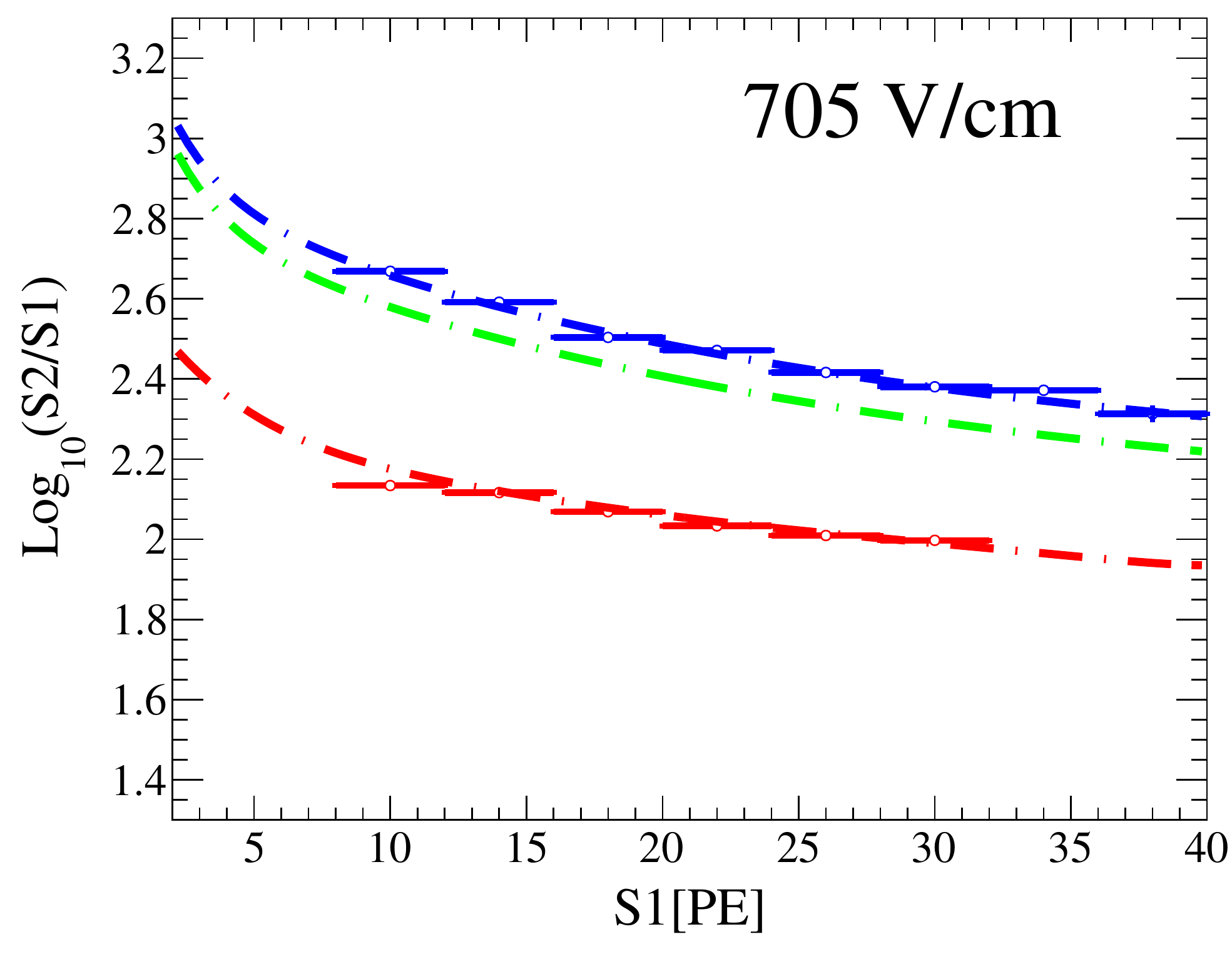}
 \includegraphics[width=5cm,height=3.75cm]{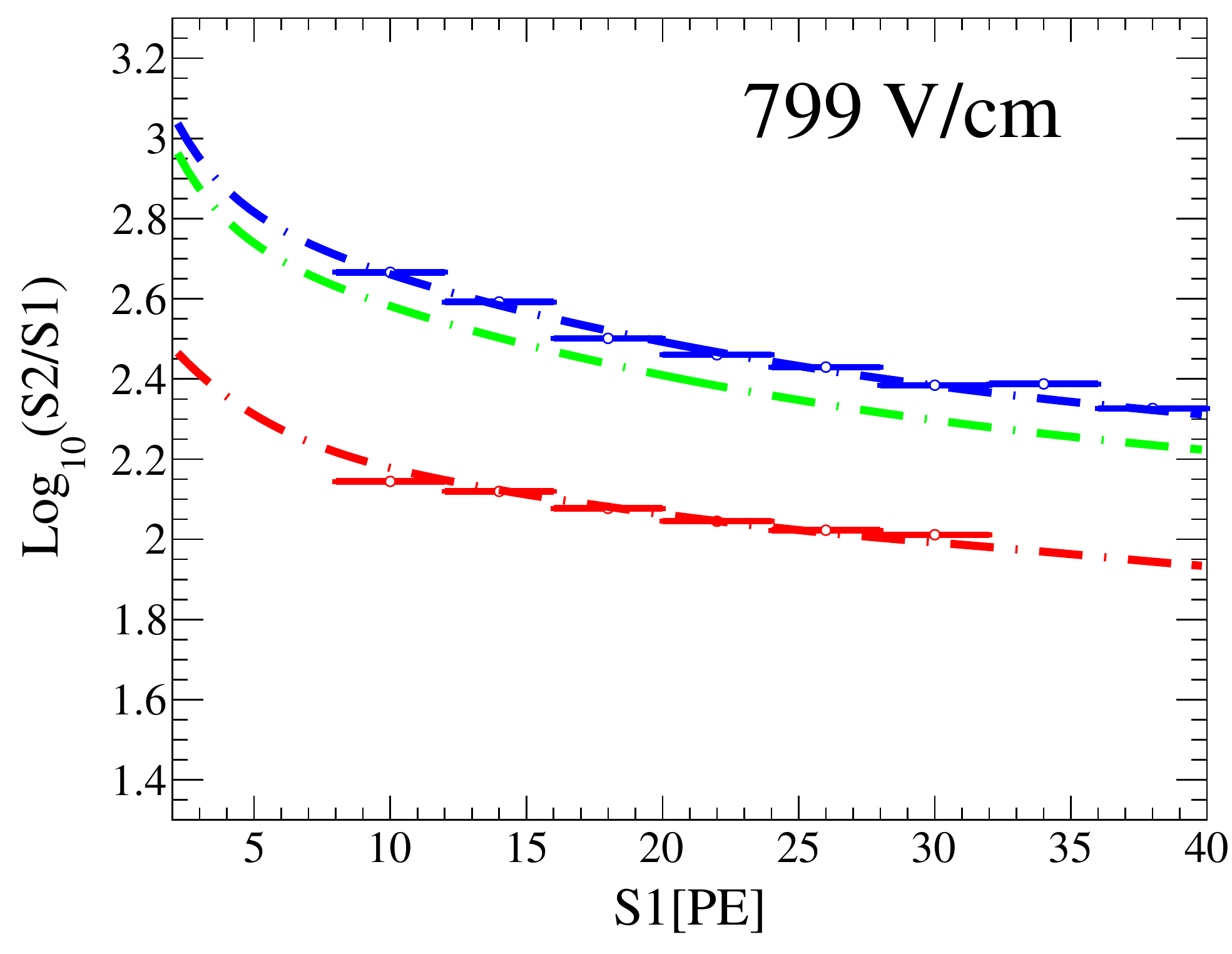}
 \includegraphics[width=5cm,height=3.75cm]{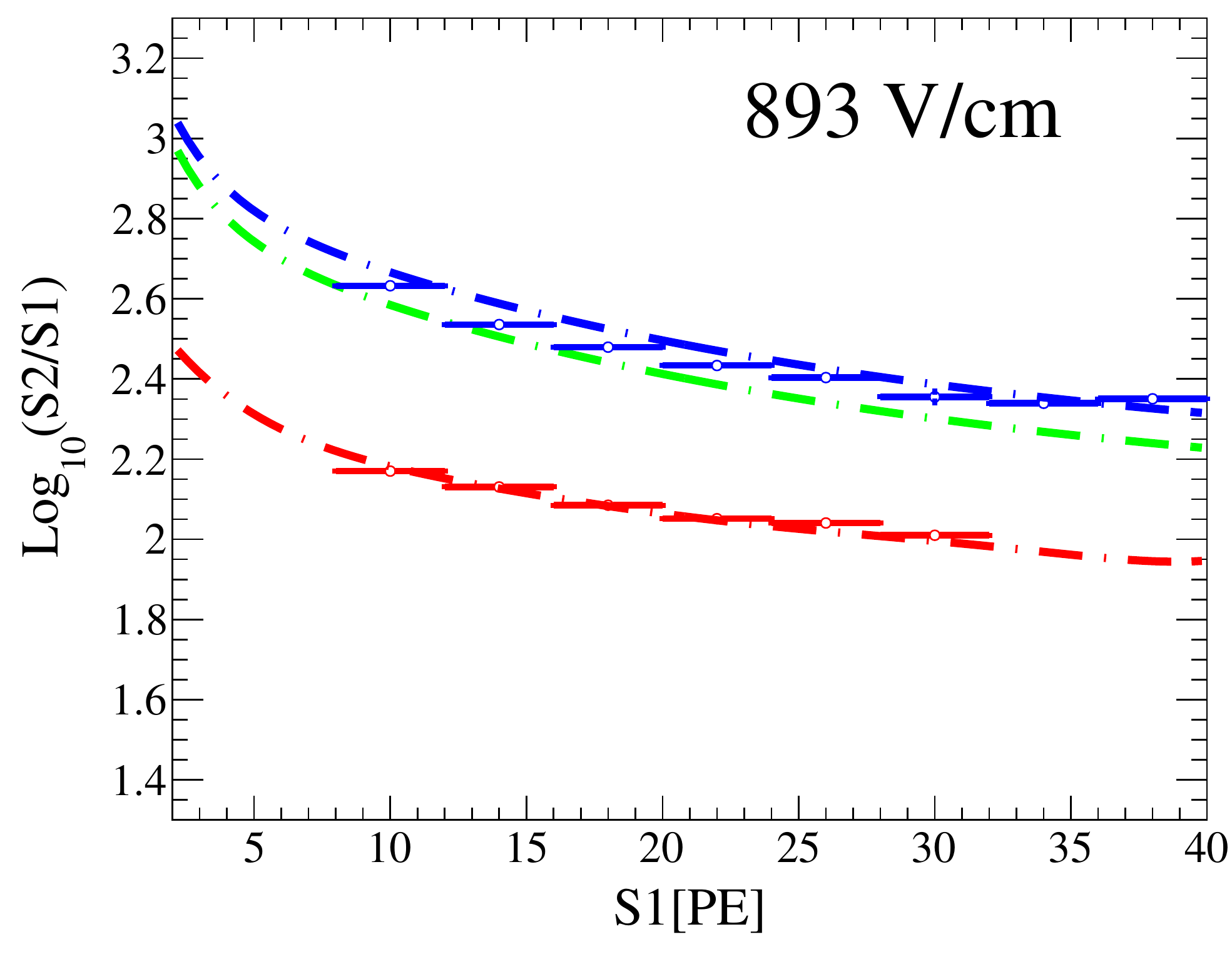}
 
 \includegraphics[width=5cm,height=3.75cm]{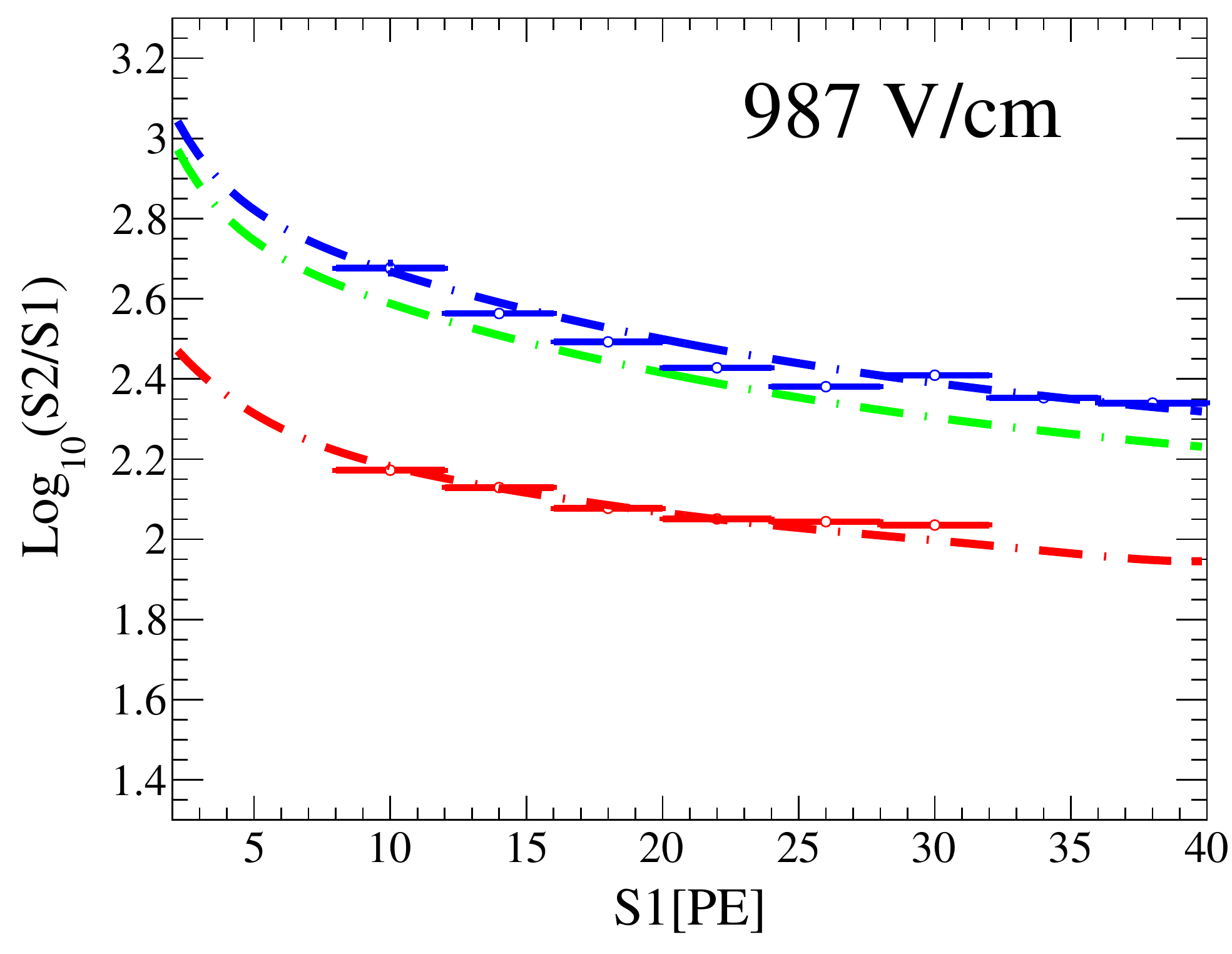}
 \includegraphics[width=5cm,height=3.75cm]{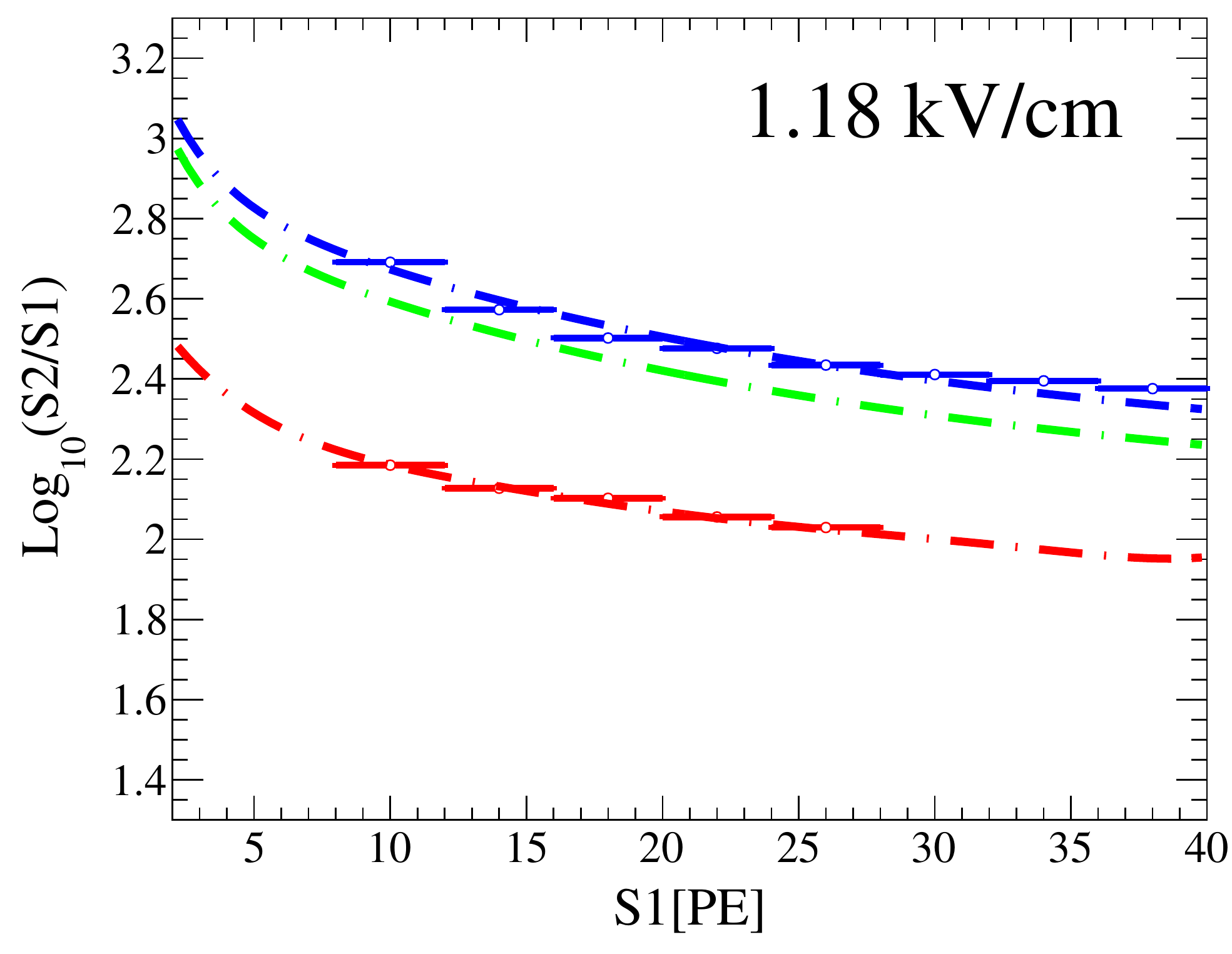}
 \includegraphics[width=5cm,height=3.75cm]{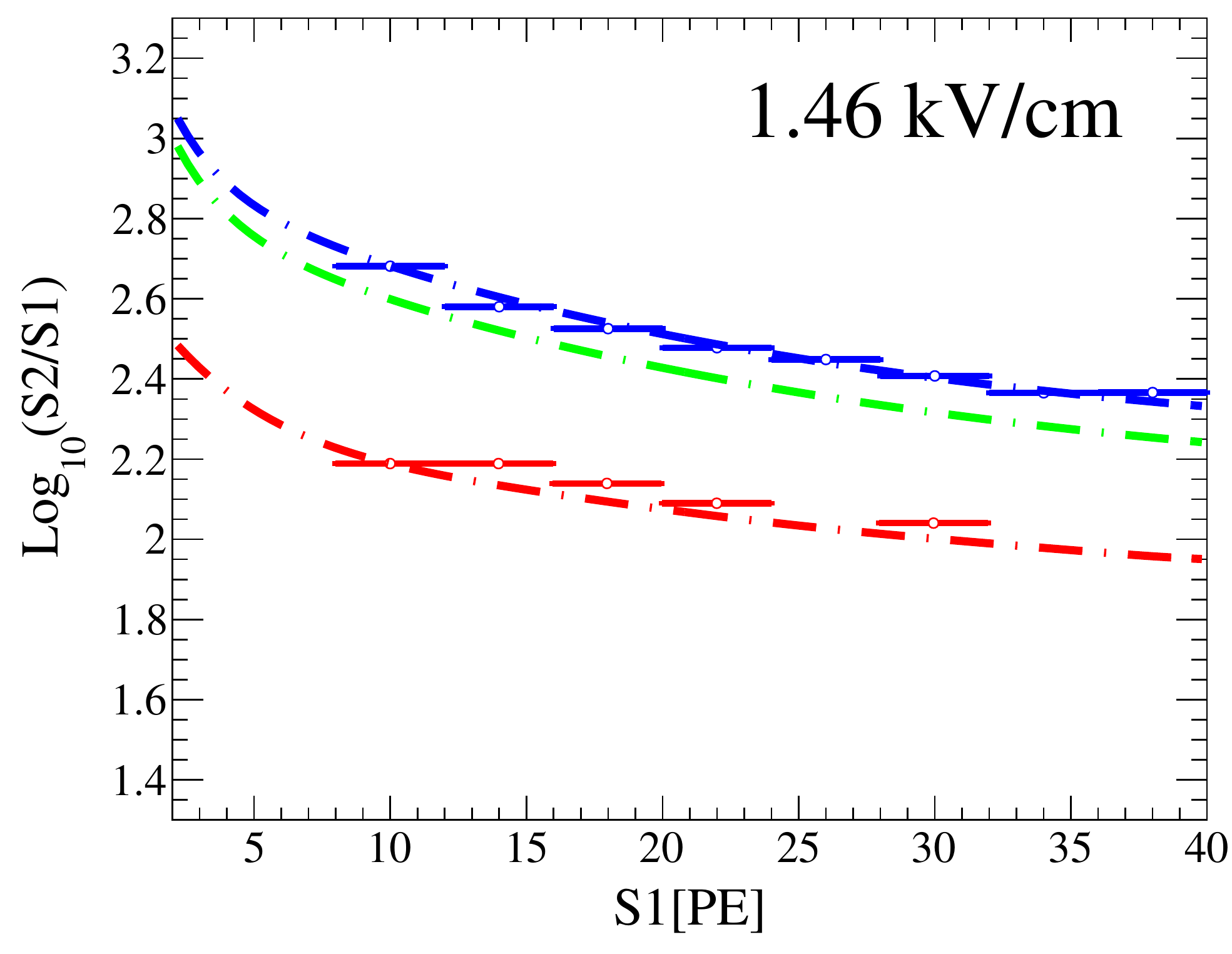}
 
 \includegraphics[width=5cm,height=3.75cm]{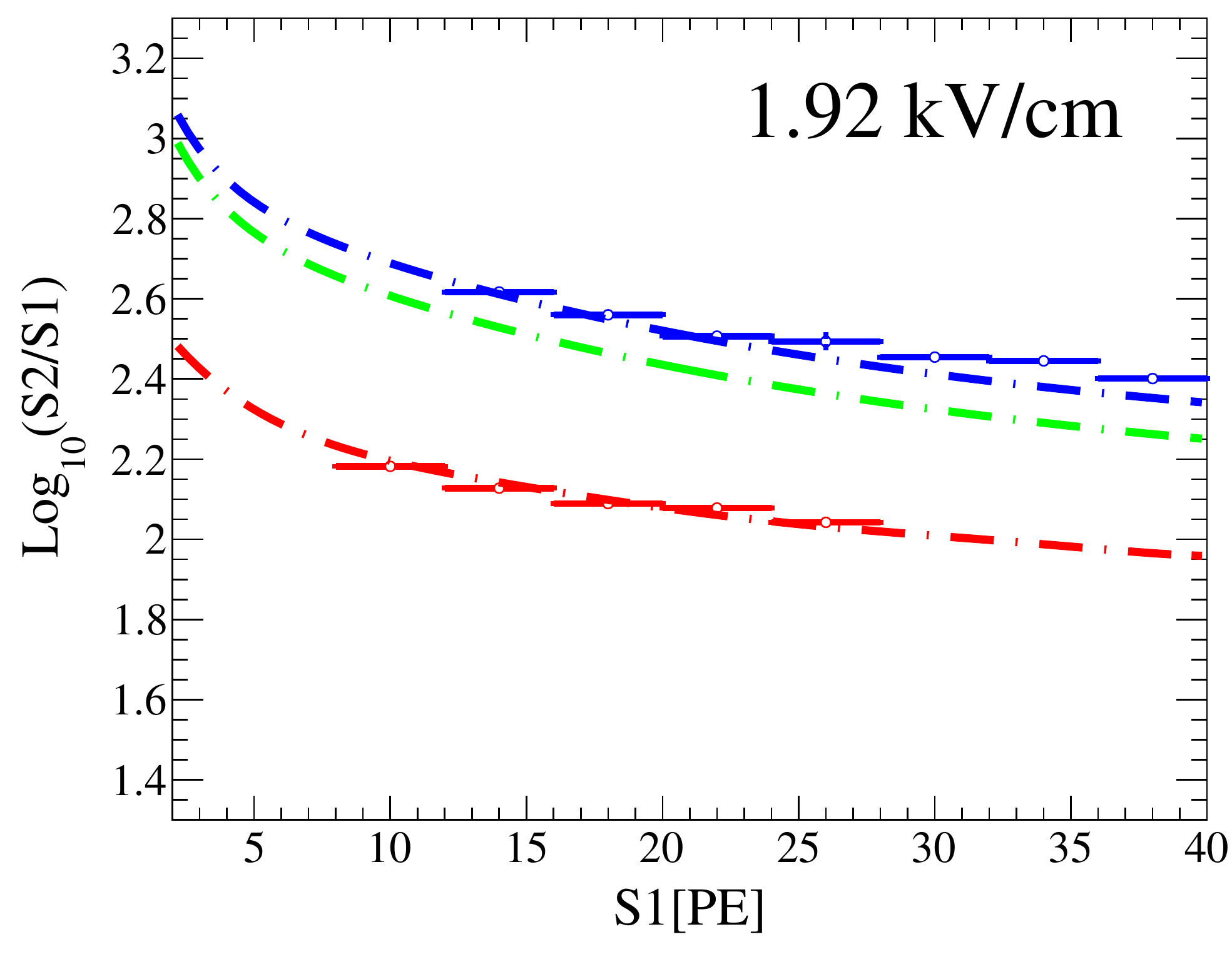}
 \includegraphics[width=5cm,height=3.75cm]{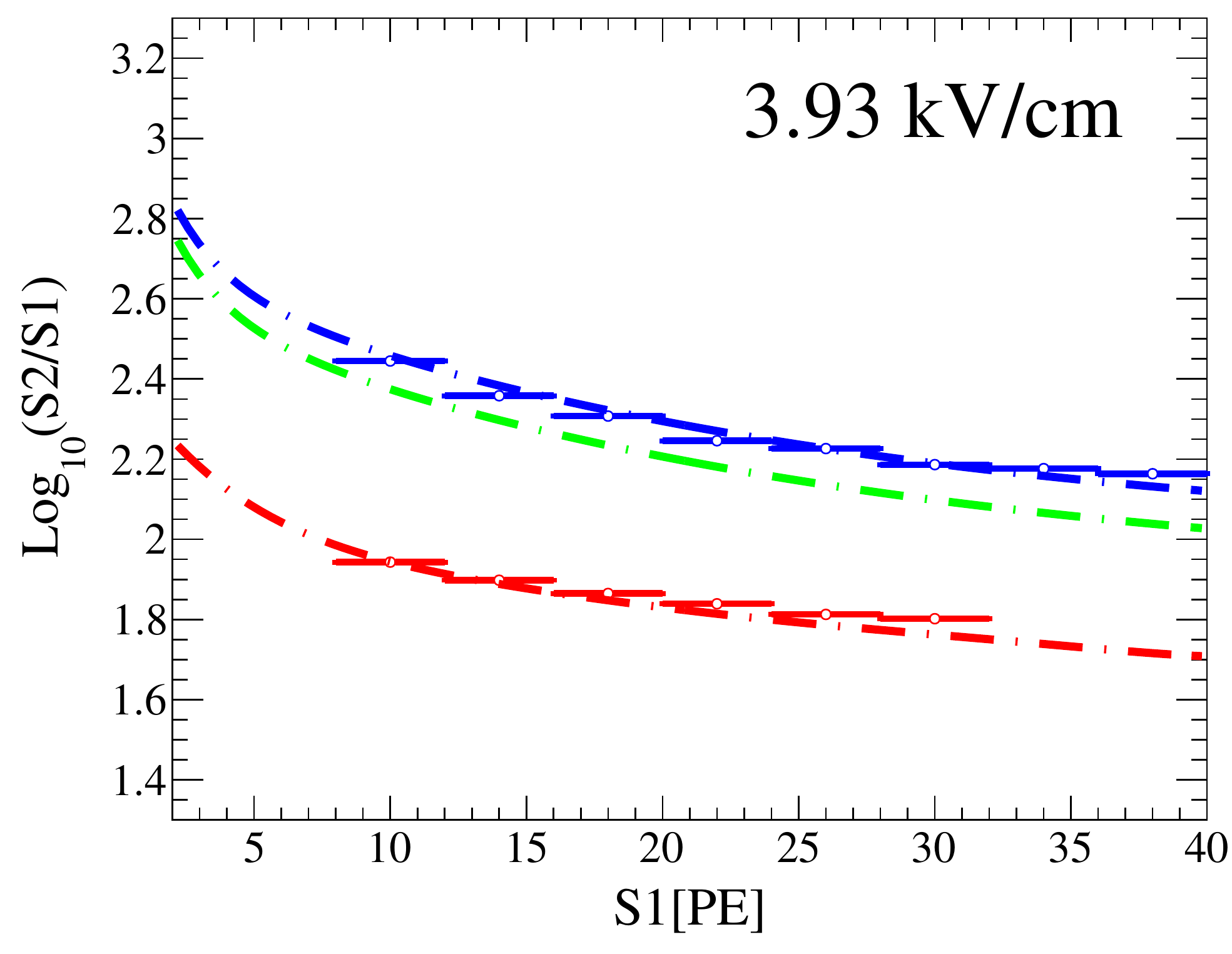} 
 \includegraphics[width=5cm,height=3.75cm]{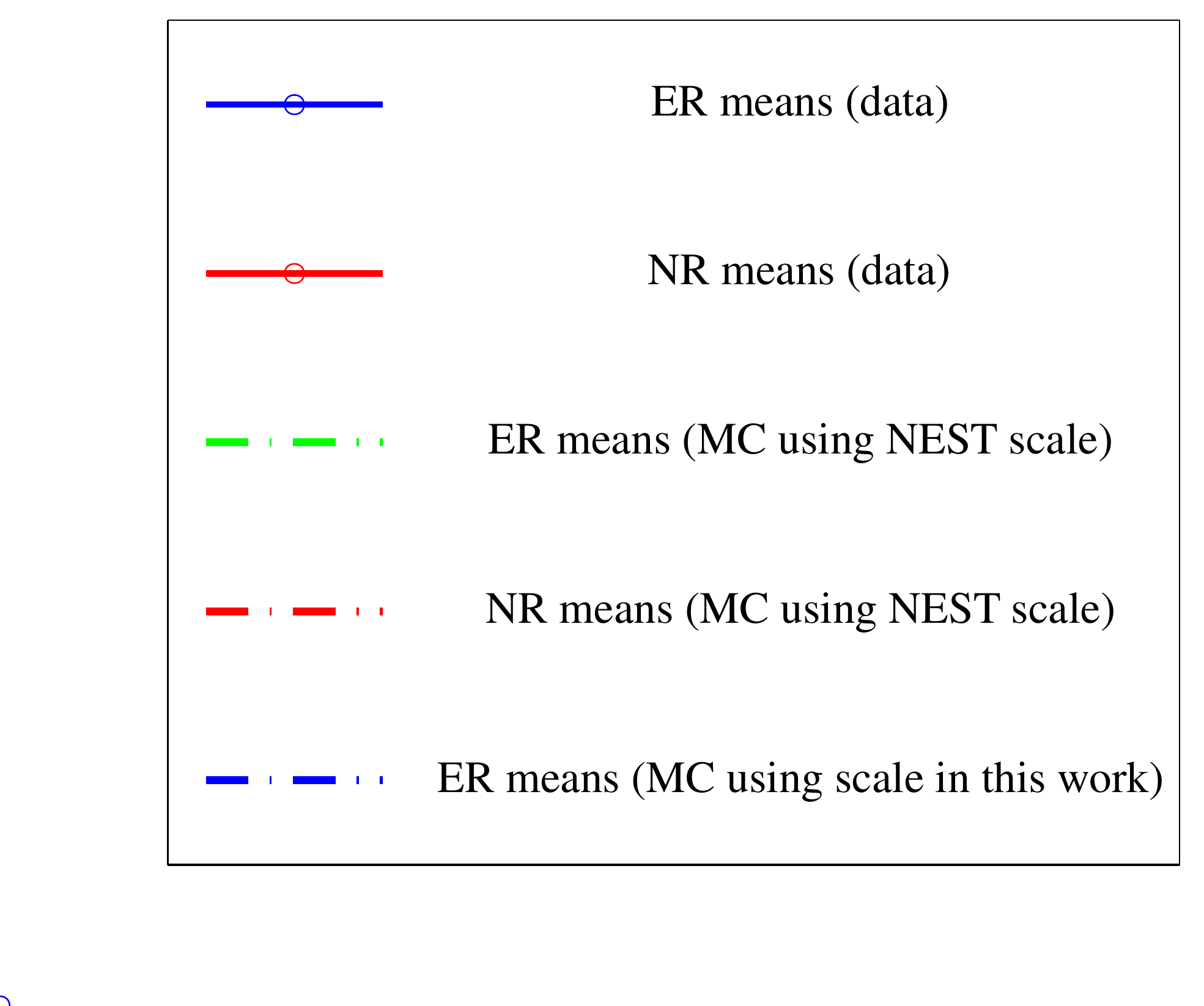}
 
 \caption{\small The measured mean values of ER (blue data points) and NR (red data points) from $Log_{10}(S2/S1)$ bands for drift fields from 236~V/cm to 3.93~kV/cm. The red dashed lines are the NR band mean lines from simulation using NEST model~\cite{NESTWeb, NEST-1.0}. The blue dashed lines are the ER band mean lines from simulation using the photon and electron response obtained in this work, as shown in Fig.~\ref{fig:ERBoxParameter}. For comparison, the ER band mean lines from simulation using the NEST model are shown as the green dashed lines, which are consistently lower than the measured values from this work. }
 \label{fig:ERNRDiffFields}
\end{figure*}

In order to extract the photon and electron yields from our measured ER and NR bands, we consider the signal generation mechanism in LXe~\cite{Aprile:RMP}.
\begin{eqnarray}
& \varepsilon = (N_\gamma+N_e) W_q & \label{eq:AntiCorrelation}\\
& N_\gamma = (N_{ex}/N_{i}+r) N_i& \label{eq:Ng}\\
& N_e = (1 - r) N_i&, \label{eq:Ne}
\end{eqnarray}
where $\varepsilon$ is the energy deposition, $W_q$ is the average energy required to produce a quanta (photon or electron). $W_q$ is found to be 13.7$\pm$0.2\,eV~\cite{NEST}, independent of energy deposition and drift field.  $N_{ex}/N_{i}$ is the ratio of number of excimers formed $N_{ex}$ to number of ions $N_i$ created from the energy deposition. $N_{ex}/N_i$ is taken as a constant at about 0.06~\cite{NEST} for ERs and is modelled as a function of the applied field and deposited energy for NRs~\cite{NESTWeb}. $r$ is the electron-ion recombination fraction. 
In the Thomas-Imel box model~\cite{BoxModel} approximation for low energy events (which is used in NEST~\cite{NEST}),  
\begin{eqnarray}
 &r = 1 - \frac{1}{\xi} \ln{(1+\xi)},& \xi = \frac{\alpha N_i}{4a^2 \mu E}. \label{eq:BoxModel}
\end{eqnarray}
Here $\mu$ and $E$ are the mobility in xenon and the field strength, respectively.  $\alpha$ and $a$ are the recombination coefficient and the box volume size~\cite{BoxModel}. The parameter $4 \xi/N_i = \alpha / ( a^2 \mu E )$ is a dimensionless constant at a given drift field~\cite{NEST, DahlThesis}.

For a given recombination fraction $r$, the photon and electron yields at a given energy can be obtained based on Eqs.~\ref{eq:AntiCorrelation}~\ref{eq:Ng}~\ref{eq:Ne}. By taking into account the detector related parameters, PDE and EAF, and the statistical effects, we simulate the ER and NR bands for comparison with the data bands, as shown in Fig.~\ref{fig:ERNRMeansComparison} (right). 
The simulation takes into account the Poisson fluctuation of photon detection and Binomial fluctuation of the electron-ion recombination, as well as a Gaussian fluctuation on the recombination fraction $\Delta$r($\varepsilon$). 
Additionally, the fluctuations caused by the PMT's single-photoelectron (SPE) resolution and the gas gain, which is not significant due to a relatively large number of electrons, are taken into account.  
Both the electron and NR energy spectra are obtained from a Geant4 simulation~\cite{G4}. 
For the ER band, the simulated spectrum below 20\,keV$_{ee}$ is used.
The ER and NR spectra from simulation are shown in Fig.~\ref{fig:G4ERNRSpec}.

A $\chi^2$ analysis to compare the measured and simulated ER band means  is carried out by scanning different 4$\xi/N_i$. The minimized $\chi^2$ value corresponds to the best-fit 4$\xi/N_i$ value (see Fig.~\ref{fig:ERBoxParameter} inset). 
The S1 range is from 8 to 40\,PE.
The 8\,PE threshold is constrained by the high dark rate observed in data.
The S1 range corresponds to the energy range of about 3 to 7\,keV$_{ee}$, based on the best-fit 4$\xi$/N$_i$ in this work.
The recombination fluctuation $\Delta r$ can be obtained by comparing the band widths between data and simulation. 
It is out of the scope of this paper and will be reported later. 
The uncertainty of $\Delta$r brings an uncertainty of $^{+0.004}_{-0.002}$ to the MC ER means, which will be taken into account in the interpretation of the uncertainties of 4$\xi$/N$_i$.

\begin{figure}[htp]
 \center
 \includegraphics[width=8cm, height=6.4cm]{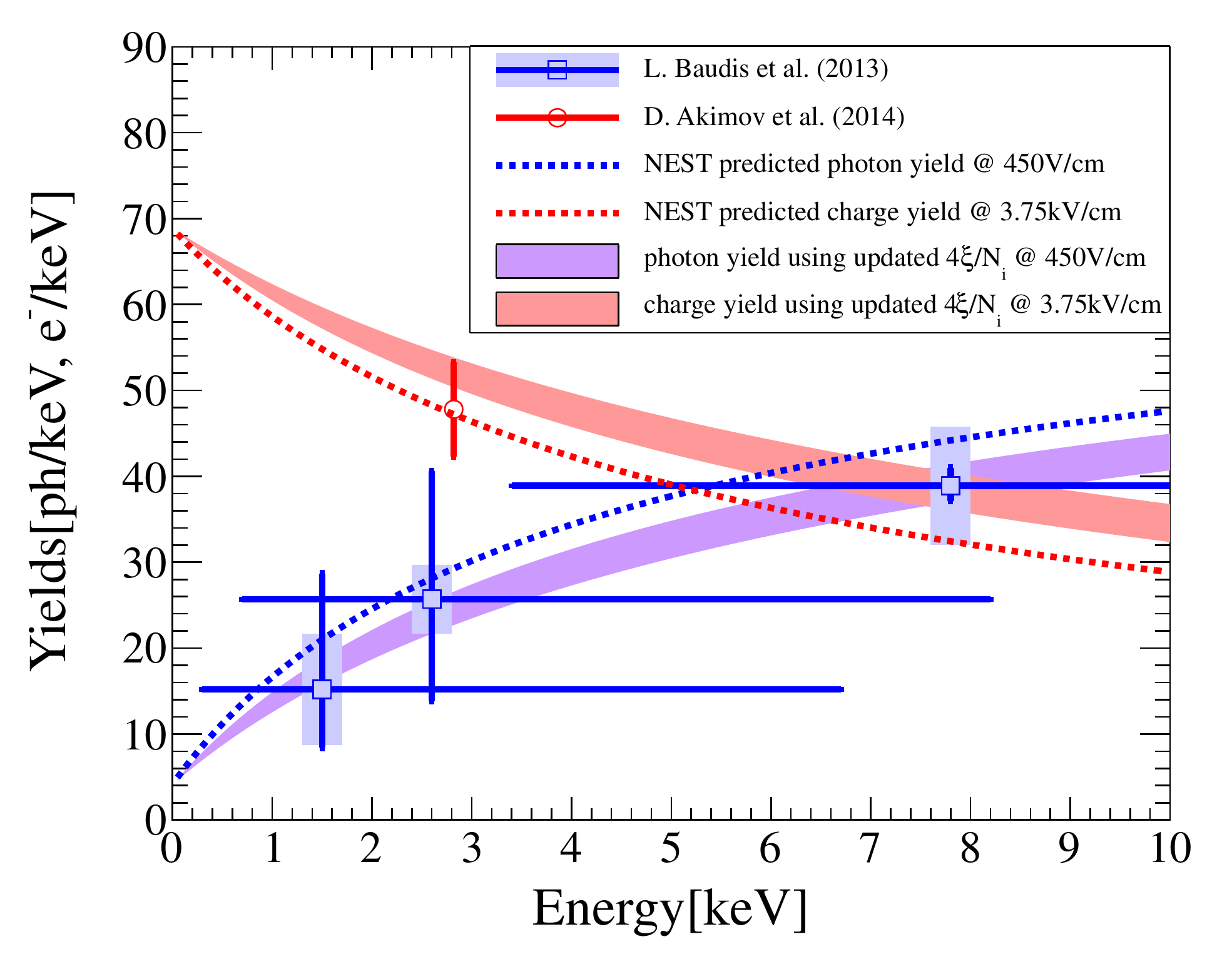}
 \caption{\small The comparison of photon and electron yields derived from our measurement and NEST to the fixed energy measurements in~\cite{LowERIonizationMeasurement} and~\cite{ZurichCompton}. The red circle is the measured ionization yield of 2.82\,keV$_{ee}$ gamma under a field of 3.75\,kV/cm~\cite{LowERIonizationMeasurement}, and the blue rectangles are the measured photon yields under a field of 450\,V/cm~\cite{ZurichCompton}. The blue shadows represent the systematic uncertainties of the photon yields from~\cite{ZurichCompton}. The statistical and systematic uncertainties of the photon yield from~\cite{ZurichCompton} are the propagations of the uncertainties of the relative yield R$_e$ and the S1 quenching under 450\,V/cm q(450).
The red (violet) shadow represents the range of the charge (photon) yields at 3.75\,kV/cm (450\,V/cm) based on the Thomas-Imel box model and the updated 4$\xi$/N$_i$ obtained in this work with the uncertainties (Fig.~\ref{fig:ERBoxParameter}).
}
 \label{fig:YieldComparison}
\end{figure}

The most probable values for 4$\xi/N_i$ obtained from the $\chi^2$ analysis for all scanned fields are shown in Fig.~\ref{fig:ERBoxParameter}. The systematic uncertainties are dominated by the uncertainties of PDE and EAF. A fit through the ER means under 236\,V/cm using Eq.~\ref{eq:BoxModel} gives a value of 0.0214$\pm$0.0003 for 4$\xi/N_i$, which is $\sim$30\% lower than the value in NEST.

Our measured $4 \xi/N_i$ values are significantly lower than those measured by Dahl~\cite{DahlThesis}. The derived predictions from NEST, which uses mainly the data from Dahl, thus give a higher value than our measurement. 
It is because that Dahl's measurement was from a LXe detector without X-Y position sensitivity and thus the edge effect gave a large systematic error.
In our measurements, the ER photon yields from data in the entire volume without radius selection are observed to be closer to the NEST predictions, although still about 2.5\,ph/keV$_{ee}$ lower than the NEST predictions. 
The ER photon yields derived from data in the central fiducial volume have larger difference, which is about 5\,ph/keV$_{ee}$ lower than the NEST predictions.

Fig.~\ref{fig:ERNRDiffFields} shows the measured ER and NR band means at all scanned fields, together with the best fit curves from simulation. For data at each field, we also plot the simulated band means using the parameters in the NEST model~\cite{NESTWeb}. The ER band means from the NEST prediction are consistently lower than from our measurement. For NRs, the band means from NEST and our measurement agree very well.

\begin{figure}[htp]
\center
\includegraphics[width=8cm, height=4cm]{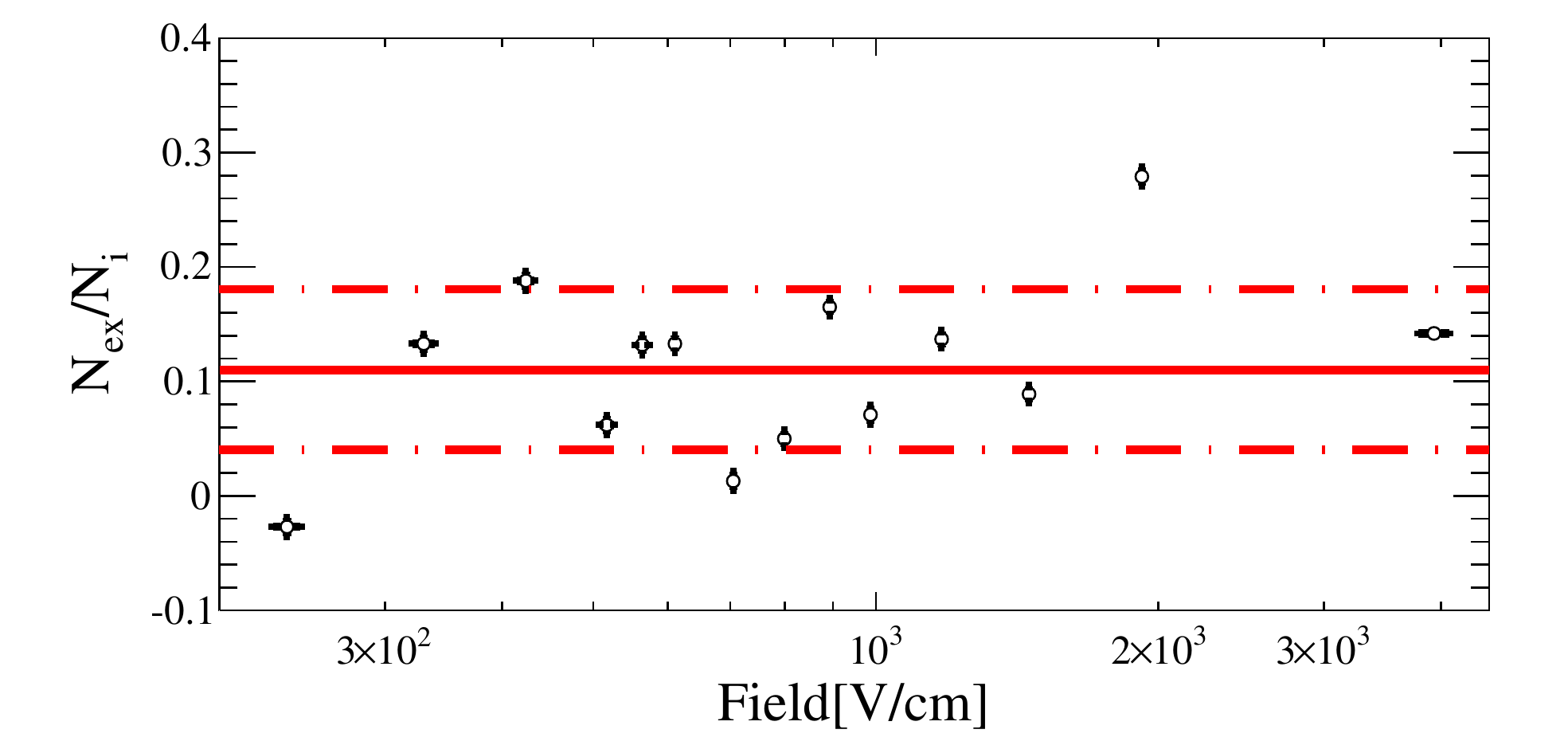}
\includegraphics[width=8cm, height=4cm]{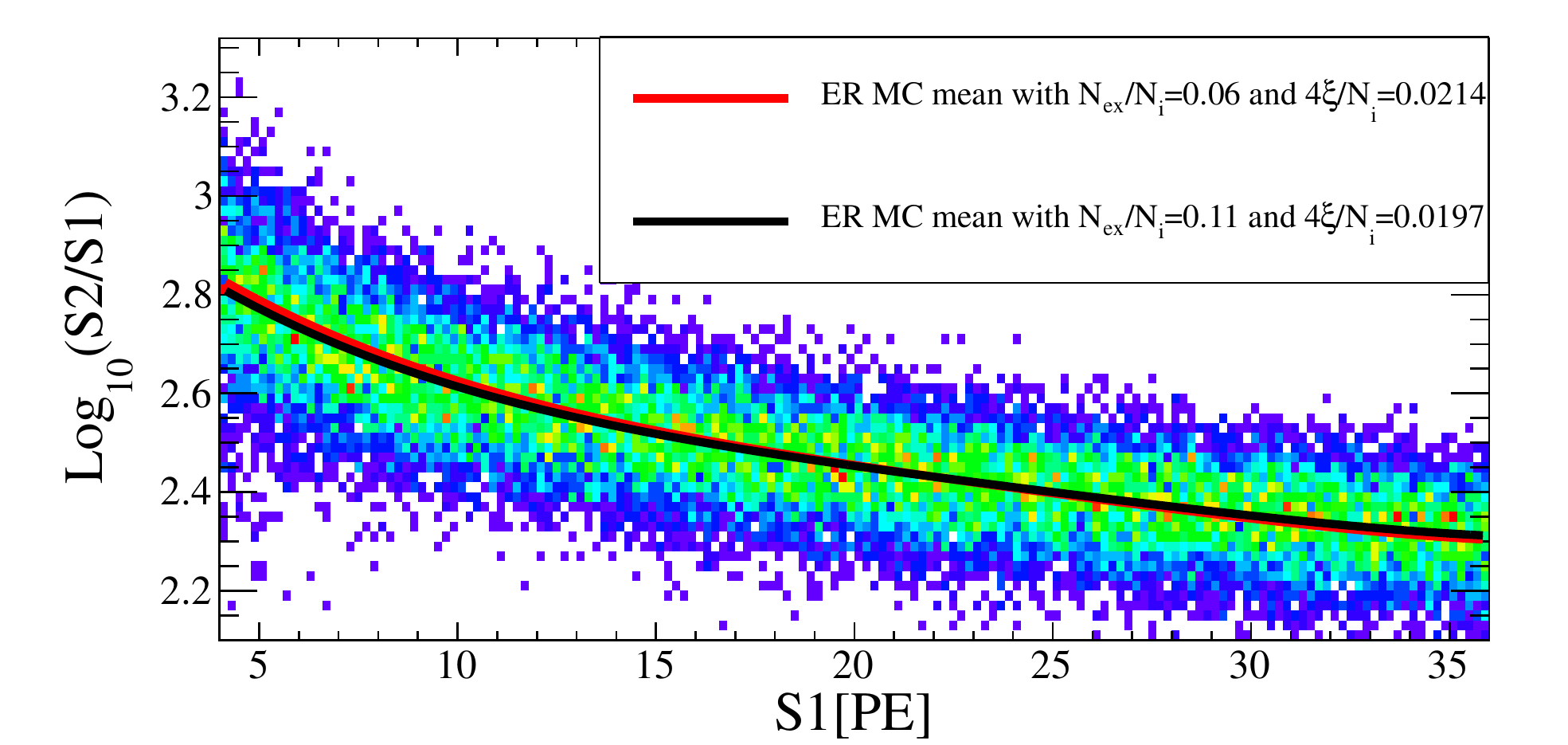}
\caption{\small (Top panel) The best-fit values of N$_{ex}$/N$_i$ obtained by comparing ER means in MC and data.
The 4$\xi$/N$_i$ and N$_{ex}$/N$_i$ are both set to be free parameters in the minimum-$\chi^2$ fitting. 
The mean N$_{ex}$/N$_i$ is 0.11 with variance of 0.07 among different fields.
With the N$_{ex}$/N$_i$=0.11, the 4$\xi$/N$_i$ obtained in this work changes to 4$\xi$/N$_i$=0.042$^{(+0.007)}_{(-0.006)}$ E$^{-0.140}$ correspondingly.
(Bottom panel) The simulated ER band with N$_{ex}$/N$_{i}$=0.11 and 4$\xi$/N$_i$ = 0.0197 (compatible with the ER measured under 236\,V/cm in this work). 
The red (black) solid line represents the mean of simulated ER band with N$_{ex}$/N$_{i}$ and 4$\xi$/N$_i$ of 0.06 (0.11) and 0.0214 (0.0197), respectively.
}
\label{fig:FloatedNexNi}
\end{figure}


\begin{figure*}[htb]
\center
\includegraphics[width=0.8\linewidth]{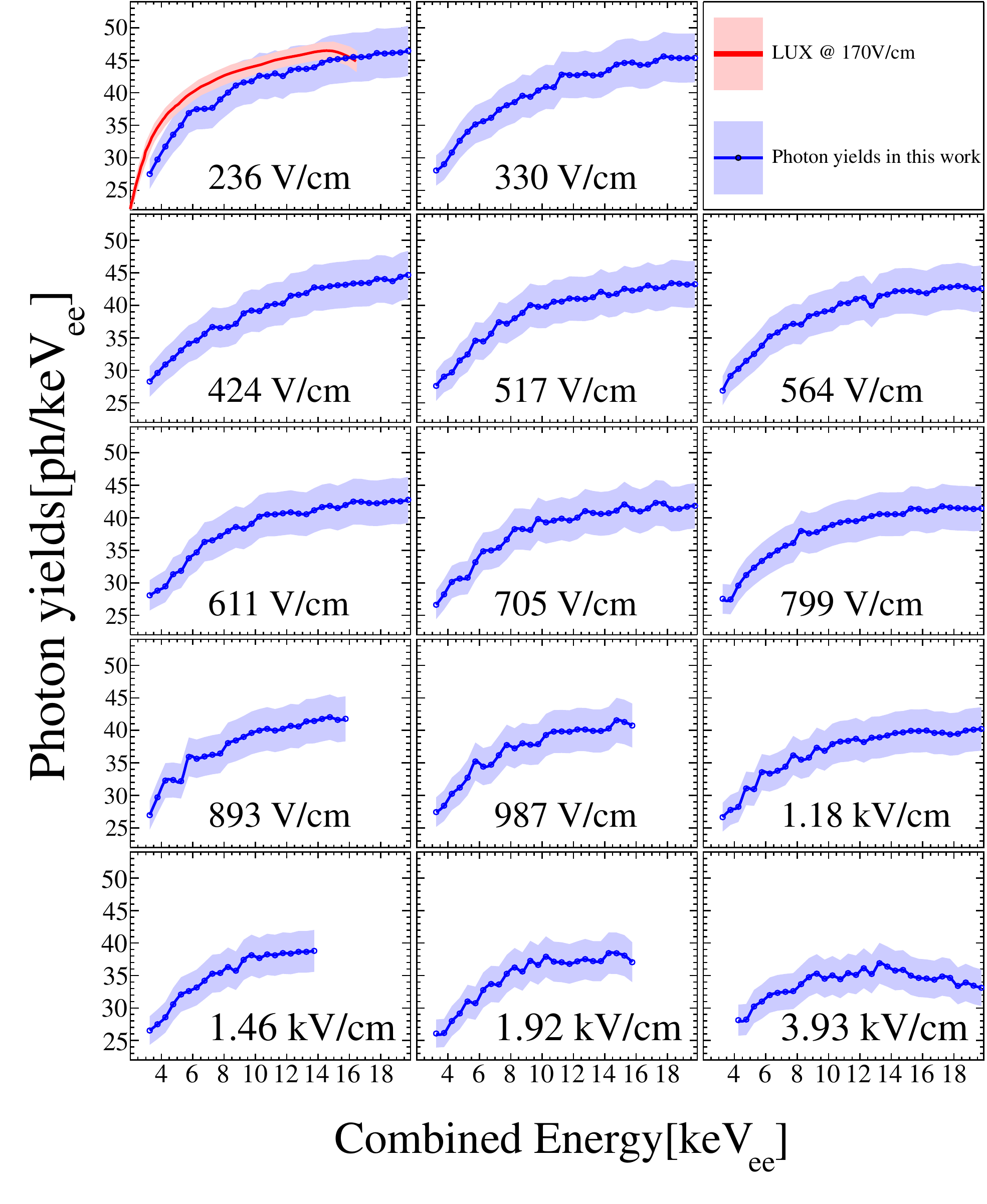}
\caption{\small The photon yields P$_y$ as a function of combined energy (blue lines) in this work at all scanned fields. The recent result from LUX~\cite{Dobi} is plotted in the left-upper panel as well.
The uncertainties (blue shadow) include those from PDE and EAF.}
\label{fig:PhotonYieldsThisWork}
\end{figure*}

Using the measured 4$\xi/N_i$ and its dependence on the field, we are able to predict the photon and electron yields at any given field from 236\,V/cm to 3.93\,kV/cm for low energy electron recoils below 7\,keV$_{ee}$. There were very few measurements of photon and electron yields with a fixed energy source. The only measured electron yield is for 2.82\,keV$_{ee}$ ERs at 3.75\,kV/cm~\cite{LowERIonizationMeasurement}. For photon yields below 10\,keV$_{ee}$ with a drift field, the only measurement is from the tagged Compton scattering experiment at 450\,V/cm~\cite{ZurichCompton}. Fig.~\ref{fig:YieldComparison} shows the predictions using 4$\xi/N_i$ values from our measurement and NEST, compared with the fixed energy measurements. For the photon yields at 1.5, 2.6, 7.8\,keV$_{ee}$ events in ~\cite{ZurichCompton}, we derive the values as the product of their relative light yield R$_e$ to  32.1\,keV$_{ee}$ events from $^{83m}$Kr at zero field, their S1 quenching q(450) under 450\,V/cm and NEST predicted photon yield at 32.1\,keV$_{ee}$, at which energy much more accurate measurements are 
available, providing more precise predictions from NEST. Due to the large uncertainties associated with the fixed energy measurement, the ER yields from our measurement and NEST are both compatible within the errors. Further measurement from other groups using the band comparison method used here or more precise measurement with fixed energy sources are needed to reconcile the differences between this result and those from NEST and Dahl.

For NRs with S1 ranging from 10 to 30\,PE (with energy approximately from 8 to 20\,keV$_{nr}$), we found that our measurement is quite consistent with the results from NEST v1.0~\cite{NEST-1.0}, thus validating the models used in NEST for predicting the response of low energy NRs for a large range of drift fields (236\,V/cm to 3.93\,kV/cm) studied in this work.

In this analysis, the excimer-to-ion ratio N$_{ex}$/N$_i$ for ER in Eq.~\ref{eq:Ng} is fixed at 0.06 (which is based on the calculation~\cite{NexNiCal1, NexNiCal2}) to be compatible with the treatment in NEST~\cite{NEST}.
However some measurements indicate a larger N$_{ex}$/N$_i$ of 0.13$\pm$0.07~\cite{DokeNexNi} and 0.20$\pm$0.13~\cite{ElenaNexNi}.
Additionally in our data, we performed a similar $\chi^2$ analysis as illustrated in Fig.~\ref{fig:ERBoxParameter}, but with both the 4$\xi$/N$_i$ and N$_{ex}$/N$_i$ treated as free parameters. 
The best-fit N$_{ex}$/N$_i$ obtained for all scanned fields are shown in Fig.~\ref{fig:FloatedNexNi}. 
The mean N$_{ex}$/N$_i$ is 0.11 with the variance of 0.07.
This is in consistent with the measurements~\cite{DokeNexNi, ElenaNexNi} and the calculation~\cite{NexNiCal1, NexNiCal2}.

According to NEST~\cite{NEST}, the Thomas-Imel box model is not valid for describing the recombination process in LXe for ER energy larger than 15\,keV$_{ee}$.
In this work, we also calculated the model-independent photon yields at all scanned fields based on the combined energy E$_c$, which is defined as:
\begin{equation}
E_c = ( \frac{S1}{PDE} + \frac{S2}{EAF} ) W_q.
\label{eq:CombinedEnergy}
\end{equation}
The mean photon yield, P$_y$, is the average number of photons generated in LXe per keV energy.
Our measured photon yields at all scanned fields are shown in Fig.~\ref{fig:PhotonYieldsThisWork}, along with the photon yields from LUX~\cite{Dobi} at 180\,V/cm.
The measured photon yields are observed to deviate from the box model when the combined energy is larger than 8\,keV$_{ee}$.
Above 8\,keV$_{ee}$, the measured photon yields are lower than the box model prediction, indicating that the Doke-Birk recombination~\cite{DokeBirk} starts to contribute to the process.
We report the results here and leave the data fitting for different recombination models at all energies for future publications, together within the NEST group.

\section{Conclusion}

In summary, we performed new measurements of scintillation and ionization of LXe for low energy electronic and NRs at drift fields from 236\,V/cm to 3.93\,kV/cm using a three-dimensional sensitive LXe time projection chamber.  The three-dimensional sensitivity allows the removal of edge events which reduces the systematic errors. The responses to NRs from our measurement are quite consistent with the NEST model. But the responses to ERs from our measurement differ from the parameters in NEST. 

By using a simulation taking into account the detector parameters and all statistical effects, we are able to reproduce the electronic and NR bands in Log$_{10}$(S2/S1) over S1 space and obtain the recombination parameters for ER at all drift fields using a minimum $\chi^2$ method. Our obtained recombination parameters for ER bands are well fit by the Thomas-Imel box model (S1 from 8 to 40\,PE), with 4$\xi/N_i$ values about 30\% lower than those in the current NEST model.

We also provide the model-independent ER photon yields as a function of the combined energy at all scanned field.
The photon yields deviate from the box model when the energy is larger than $\sim$8\,keV$_{ee}$.
These data are useful in the global analysis and modeling of the ER recombination in the low energy region.

Due to the lack of precise measurement at fixed energy below 10\,keV$_{ee}$ previously, our new measurements provide a set of best available data to be used to predict the response of low energy ERs in LXe. In addition to that, we measured the response from very low to high drift fields, providing useful data to predict the response of LXe at different drift fields for future large LXe dark matter detectors.

\begin{acknowledgments}
The authors would like to thank Matthew Szydagis for valuable comments during the preparation of this paper.
This work is supported by National Science Foundation of China (Grant No.: 11375114 and 11175117), and Science and Technology Commission of Shanghai Municipality (Grant No.: 11DZ2260700).
\end{acknowledgments}

\input bibliography.tex
\end{document}

%% file: author_list.tex
\author{Qing~Lin}
\author{Jialing~Fei}
\author{Fei~Gao}
\author{Jie~Hu}
\email[Now at Department of Physics, University of Alberta, Canada]{}
\author{Yuehuan~Wei}
\email[Now at Institute of Physics, University of Z\"urich, Switzerland]{}
\author{Xiang~Xiao}
\affiliation{Department of Physics and Astronomy, Shanghai Key Laboratory for Particle Physics and Cosmology, Shanghai Jiao Tong University, Shanghai, 200240, China}
\author{Hongwei~Wang}
\affiliation{Institute of Applied Physics, Chinese Academy of Sciences, Shanghai, China}
\author{Kaixuan~Ni}
\email[Corresponding author: ]{nikx@physics.ucsd.edu}
\affiliation{Department of Physics, University of California, San Diego, CA, USA}